\documentclass[twocolumn,astrosymb]{aastex631}

\usepackage{xcolor}
\usepackage{mathtools}
\usepackage{amsmath,amsthm,amssymb,amsfonts,bm}
\usepackage{soul}
\usepackage[super]{nth}
\usepackage{textcomp}
\usepackage{url}

\newcommand{\rxj}{RX~J1347.5-1145}
\newcommand{\rtwofive}{R$_{2500}$}
\newcommand{\ytwofive}{$\langle y \rangle_{2500}$}
\newcommand{\Ttwofive}{$\langle \textrm{T}_{\textrm{sz}} \rangle_{2500}$}

\newcommand{\Tpwxtwofive}{$\langle \textrm{T}_{\textrm{x,pw}}\rangle_{2500}$}

\newcommand{\ASZpsw}{$\textrm{A}_{\textrm{PSW}}^{\textrm{SZ}}$}
\newcommand{\ASZpmw}{$\textrm{A}_{\textrm{PMW}}^{\textrm{SZ}}$}
\newcommand{\ASZplw}{$\textrm{A}_{\textrm{PLW}}^{\textrm{SZ}}$}
\newcommand{\ASZb}{$\textrm{A}_{b}^{\textrm{SZ}}$}
\newcommand{\nsrc}{N$_{\textrm{src}}$}
\newcommand{\dIbolo}{$\langle d\textrm{I}_{\textrm{Bcam}} \rangle_{2500}$}
\newcommand{\dIplw}{$\langle d\textrm{I}_{\textrm{PLW}} \rangle_{2500}$}
\newcommand{\dIpmw}{$\langle d\textrm{I}_{\textrm{PMW}} \rangle_{2500}$}
\newcommand{\dIpsw}{$\langle d\textrm{I}_{\textrm{PSW}} \rangle_{2500}$}
\newcommand{\dIb}{$\langle d\textrm{I}_{b} \rangle_{2500}$}
\newcommand{\Bzero}{$B_{0}^b$}

\graphicspath{{./}{figures/}}

\begin{document}

\title{Measurement of the Relativistic Sunyaev-Zeldovich Correction in RX J1347.5-1145}


\author[0000-0002-0941-0407]{Victoria L. Butler}
\altaffiliation{Equal contribution.}
\email{vlb9398@rit.edu}
\affiliation{Rochester Institute of Technology, 1 Lomb Memorial Drive, Rochester, NY 14623, USA}
\author[0000-0002-9330-8738]{Richard M. Feder}
\altaffiliation{Equal contribution.}
\affiliation{California Institute of Technology Division of Physics, Math, and Astronomy, 1200 East California Boulevard, Pasadena, CA 91125, USA}

\author[0000-0002-6939-9211]{Tansu Daylan}
\affiliation{Kavli Institute for Astrophysics and Space Research, Massachusetts Institute of Technology, Cambridge, MA 02139}
\affiliation{Department of Astrophysical Sciences, Princeton University, 4 Ivy Lane, Princeton, NJ 08544}

\author[0000-0002-8031-1217]{Adam B. Mantz} 
\affiliation{Kavli Institute for Particle Astrophysics and Cosmology, Stanford University, 452 Lomita Mall, Stanford, CA 94305, USA}

\author[0000-0002-7485-7563]{Dale Mercado} 
\affiliation{Rochester Institute of Technology, 1 Lomb Memorial Drive, Rochester, NY 14623, USA} 

\author[0000-0003-4229-381X]{Alfredo Monta\~na} 
\affiliation{$^{1}$Consejo Nacional de Ciencia y Tecnolog\'ia, Av. Insurgentes Sur 1582, Col. Cr\'edito Constructor, Alcald\'ia Benito Ju\'arez, C.P. 03940, Ciudad de M\'exico, M\'exico\\$^{2}$Instituto Nacional de Astrof\'isica \'Optica y Electr\'onica, Luis Enrique Erro 1, CP 72840, Tonantzintla, Puebla, M\'exico\\} 

\author[0000-0001-8132-8056]{Stephen K. N. Portillo}
\affiliation{DIRAC Institute, Department of Astronomy, University of Washington, 3910 15th Ave. NE, Seattle, WA 98195 USA}

\author[0000-0002-8213-3784]{Jack Sayers} 
\affiliation{California Institute of Technology Division of Physics, Math, and Astronomy, 1200 East California Boulevard, Pasadena, CA 91125, USA} 

\author[0000-0002-9813-0270]{Benjamin J. Vaughan} 
\affiliation{Rochester Institute of Technology, 1 Lomb Memorial Drive, Rochester, NY 14623, USA} 

\author[0000-0001-8253-1451]{Michael Zemcov} 
\affiliation{Rochester Institute of Technology, 1 Lomb Memorial Drive, Rochester, NY 14623, USA}
\affiliation{Jet Propulsion Laboratory, 4800 Oak Grove Drive, Pasadena, CA 91109, USA}

\author[0000-0002-0350-4488]{Adi Zitrin}
\affiliation{Physics Department, Ben-Gurion University of the Negev, P.O. Box 653, Be'er-Sheva 84105, Israel}

\vspace{1.5cm}
\begin{abstract}

We present a measurement of the relativistic corrections to the thermal Sunyaev-Zel'dovich (SZ) effect spectrum, the rSZ effect, toward the massive galaxy cluster \rxj\ by combining sub-mm images from {\it Herschel}-SPIRE with mm-wave Bolocam maps. Our analysis simultaneously models the SZ effect signal, the population of cosmic infrared background (CIB) galaxies, and galactic cirrus dust emission in a manner that fully accounts for their spatial and frequency-dependent correlations. Gravitational lensing of background galaxies by \rxj\ is included in our methodology based on a mass model derived from {\it HST} observations.
Utilizing a set of realistic mock observations, we employ a forward modelling approach that accounts for the non-Gaussian covariances between observed astrophysical components to determine the posterior distribution of SZ effect brightness values consistent with the observed data. We determine a maximum \textit{a posteriori} (MAP) value of the average Comptonization parameter of the intra-cluster medium (ICM) within \rtwofive\ to be \ytwofive\ $= 1.56 \times 10^{-4}$, with corresponding 68~per cent credible interval $[1.42,1.63] \times 10^{-4}$, and a MAP ICM electron temperature of \Ttwofive\ $= 22.4$~keV with 68~per cent credible interval spanning $[10.4,33.0]$~keV.  This is in good agreement with the pressure-weighted temperature obtained from {\it Chandra} X-ray observations, \Tpwxtwofive\ $= 17.4 \pm 2.3$~keV. We aim to apply this methodology to comparable existing data for a sample of 39 galaxy clusters, with an estimated uncertainty on the ensemble mean \Ttwofive\ at the $\simeq 1$~keV level, sufficiently precise to probe ICM physics and to inform X-ray temperature calibration.
\end{abstract}


\section{Introduction} 
\label{sec:intro}

Galaxy clusters are the largest bound objects in the Universe, and gravity is the dominant process driving their evolution and setting their overall physical characteristics. For our analysis, these objects are largely self-similar in their properties \citep{Kaiser1986}. However, deviations from self-similarity can occur from processes outside of simple matter aggregation, such as major mergers between similar mass halos, feedback from active galactic nuclei (AGN), and gas sloshing within the cluster potential \citep{Zhao2003,McNamara2007,Markevitch2007}. In some cases, such as cosmological studies relying on precise halo mass estimates, it is important to quantify and understand the effects of these more complicated processes \citep{Pratt_2019}. Such deviations from self-similarity are particularly evident in the spatially resolved thermodynamics of the gaseous intra-cluster medium (ICM), which have typically been studied via the thermal Bremsstrahlung emission in X-rays \citep{Sarazin1986}. However, these studies are generally more difficult and/or impractical at higher redshifts and larger cluster-centric radii, due to $(1+z)^{4}$ cosmological dimming and the density squared dependence of the X-ray surface brightness \citep{McDonald2017,Reiprich2013}. In addition, the X-ray instruments on board {\it Chandra} and XMM-{\it Newton} have limited sensitivity to ICM gas above $\sim 10$~keV, and such temperatures are expected to be relatively common among the population of massive galaxy clusters, for example, in the shock-heated gas that results from major mergers \citep{Markevitch2002}.

\begin{figure}
    \centering
    \includegraphics[width=0.47\textwidth]{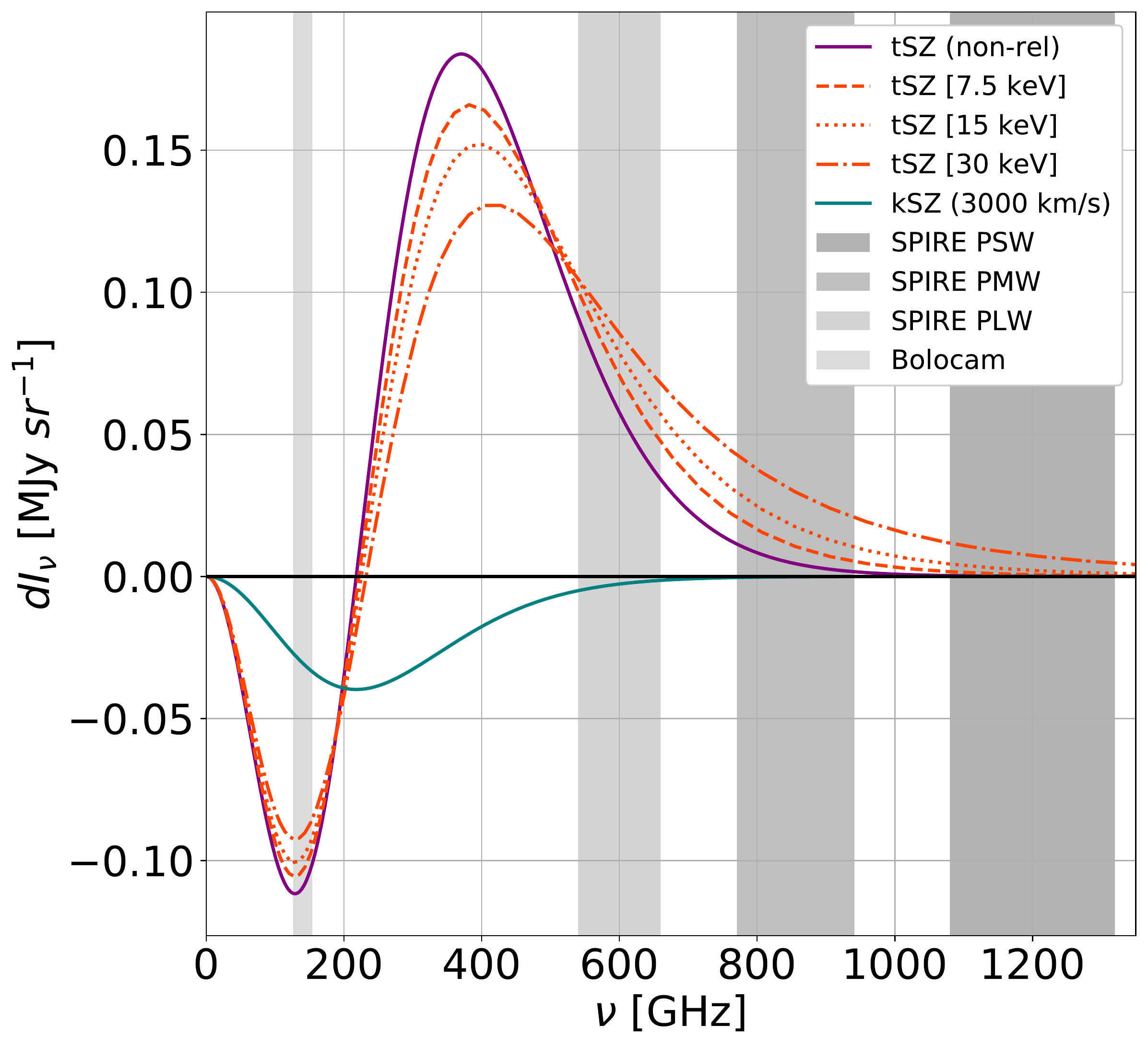}
    \caption{The thermal SZ effect spectrum with and without relativistic corrections, along with kSZ effect. The thermal SZ effect spectra were calculated assuming a total Comptonization parameter $y= 10^{-4}$ and four different temperature values. As temperature increases, the overall impact of the rSZ effect is to shift the thermal SZ effect spectrum to higher frequencies. The kSZ effect was determined using the same value of $y$ combined with a peculiar velocity of $+3000$ km s$^{-1}$ and an electron temperature of 17 keV.}
    \label{fig:3_sze}
\end{figure}

A complementary method to study ICM thermodynamics is through the Sunyaev-Zeldovich (SZ) effect signal, which is due to inverse Compton scattering of Cosmic Microwave Background (CMB) photons with energetic electrons in the ICM \citep{Sunyaev1972}. The electron population has both random thermal motions and coherent velocities, the latter due to internal turbulence or the overall peculiar velocity of the cluster. These two properties of the ICM give rise to a thermal and a kinematic SZ effect. In addition, the ICM is mildly relativistic, with an average normalized temperature equal to approximately 1\% of the normalized electron mass. This gives rise to generally mild temperature-dependent relativistic corrections to the thermal SZ effect spectrum known as the rSZ effect signal \citep[][]{Wright1979,Itoh1998,Chluba2012,Chluba2013}. The typical signatures of these signals are shown in Figure \ref{fig:3_sze}. These relativistic corrections decrease the magnitude of the SZ effect signal at $\lesssim 500$ GHz, while boosting the signal at higher frequencies.

While X-ray measurements are a more mature probe, with current satellites providing spectroscopic imaging with orders of magnitude better statistics and angular resolution than typical SZ effect observations \citep[][]{Fabian2011}, the SZ effect can play a critical role in several important areas of study. First, because the SZ effect is a fractional distortion of the CMB, the surface brightness is independent of redshift, and so it can more uniformly probe objects across cosmic time \citep[][]{Bleem2015}. In addition, since the SZ effect signal is due to photon-electron scattering rather than the ion-electron acceleration that sources the X-ray Bremmstrahlung, it is relatively brighter in the lower density cluster outskirt regions \citep[][]{Planck2013_V}. Finally, the rSZ effect signal is sensitive to arbitrarily large temperatures above the band limitations of facilities such as {\it Chandra} and XMM-{\it Newton}. 

While there are numerous examples of SZ effect studies pushing to new regimes at high redshift and/or large cluster-centric radii, relatively little effort has been directed toward measuring ICM temperatures with the rSZ effect. This is due to the combination of sensitivity needed to measure the rSZ effect signal and the myriad spatial and spectral data required to separate contamination from other astronomical sources. For example, the contrast between the rSZ effect signal and the canonical tSZ effect signal is maximized at higher frequencies in the sub-millimeter, where the total extragalactic signal is dominated by the Cosmic Infrared Background \citep[CIB,][]{Hauser1998}. 

The difficulty in measuring the rSZ effect is evidenced by the relatively modest constraints obtained to date. The first measurement to indicate a deviation from the non-relativistic tSZ effect spectrum (at $\simeq 2\sigma$) was reported by \citet{Zemcov2010} based on SPIRE observations of the Bullet cluster \citep[see also][]{Prokhorov2011,Prokhorov2012}. Subsequent observations using the Z-Spec spectrometer obtained a similar detection significance of the rSZ effect in \rxj\ \citep{Zemcov12}. 

Recent attempts to measure relativistic SZ effect corrections have been based on stacking hundreds of galaxy clusters from the {\it Planck} all sky survey \citep{Hurier2016,Erler18}, with these analyses measuring the rSZ effect signal with a statistical significance similar to what was achieved in the earlier single-object studies.  Looking forward, planned and potential sub-mm/mm facilities, like LMT/TolTEC \citep{Hughes2020,Wilson2020}, the upcoming Fred Young Sub-mm Telescope (FYST; formerly CCAT-prime) \citep{CCAT2021}, the potential Atacama Large Aperture Sub-mm/mm Telescope (AtLAST; \citealt{atLAST, atLAST_2}) and the Chajnantor Sub/millimeter Survey Telescope (CSST; \citealt{Padin2014}) hold the promise of delivering the required sensitivity, angular resolution, and field of view to make high-significance rSZ effect measurements routine. 

Here we report results from a measurement of the rSZ effect signal toward the galaxy cluster \rxj\ using data from \textit{Herschel}-SPIRE, \textit{Bolocam}, \textit{Planck}, \textit{Chandra} and the {\it Hubble Space Telescope} ({\it HST}). \rxj\ is a famous system that has been the subject of numerous SZ effect studies over the past two decades \citep{Pointecouteau1999, Komatsu2001, Pointecouteau2001, Kitayama2004, Zemcov2012, Sayers2016_RXJ, Kitayama2016}. It is one of the most massive galaxy clusters observed, with a well-defined cool core and a highly relaxed morphology over most of its projected surface \citep{Allen2002, Gitti2004, Mantz2015a}. However, there is clear evidence for a shock to the SE of the core, with a corresponding enhancement to the thermal SZ effect signal in that region \citep{Mason2010, Plagge2013, Ueda2018, DiMascolo2019}. Detailed dynamical analyses of \rxj\ indicate that this shock is the result of a 10-to-1 mass ratio merger occurring primarily in a direction orthogonal to the line of sight \citep{Johnson2012}. While this merger has produced sloshing features in the core gas, it has not significantly disrupted the cool core nor the overall relaxed morphology of the system. Furthermore, the SZ effect enhancement is relatively small in both amplitude and angular extent compared to the bulk signal and, as shown in \citet{Sayers2016_RXJ}, the enhancement is not evident in the arcminute-resolution imaging we employ in this analysis.

This work has resulted in the development of several novel techniques to address the challenges of separating unwanted astrophysical contaminants from the rSZ effect signal. The reduction of the raw data products is described in Section \ref{sec:data_reduc}, and our map fitting procedure is outlined in Section \ref{sec:map_fit}. A mock observation pipeline is required to accurately characterize our results, and the details of this pipeline are described in Section \ref{sec:mock_pipe}. Our analysis of the mocks to assess the uncertainties and biases in our measured SZ effect brightness values is reported in Section \ref{sec:error}. We present the procedure used to fit for the average Comptonization parameter and ICM temperature in Section \ref{sec:spec_fit}. Finally, we discuss the implications of this work for current and future studies in Section \ref{sec:results}. Throughout this work, we assume a flat $\Lambda$CDM cosmology with $\Omega_{\rm m} = 0.3$, $\Omega_{\Lambda} = 0.7,$ and $h = 0.7$.

\section{Data and Reduction} \label{sec:data_reduc}

    \begin{figure}
      \centering
      \includegraphics[width=0.38\textwidth]{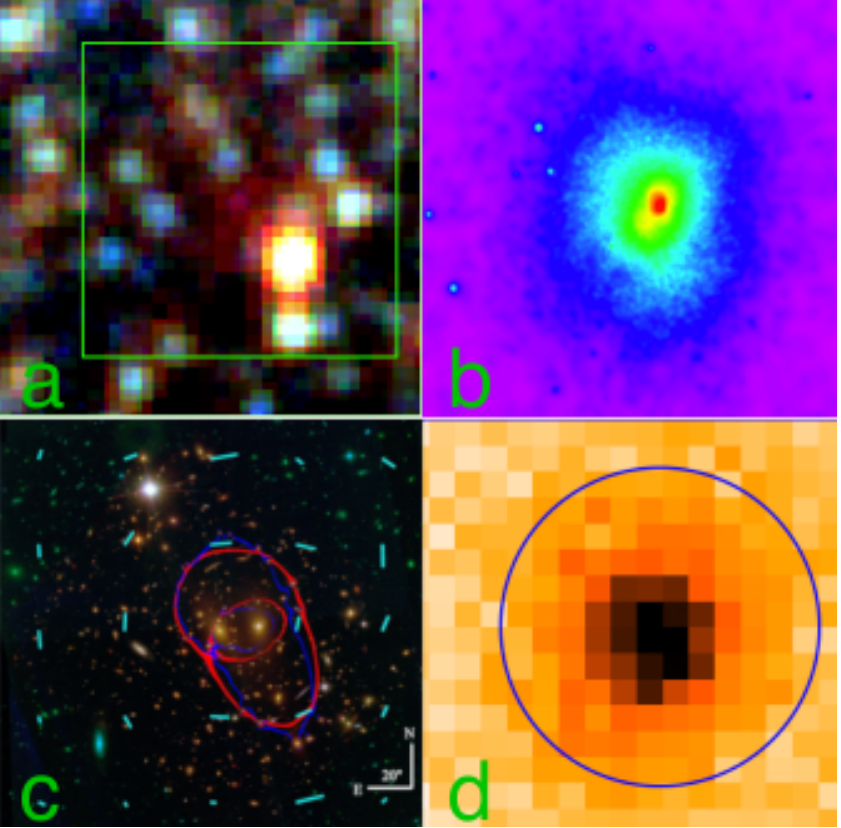}
      \includegraphics[width=0.38\textwidth]{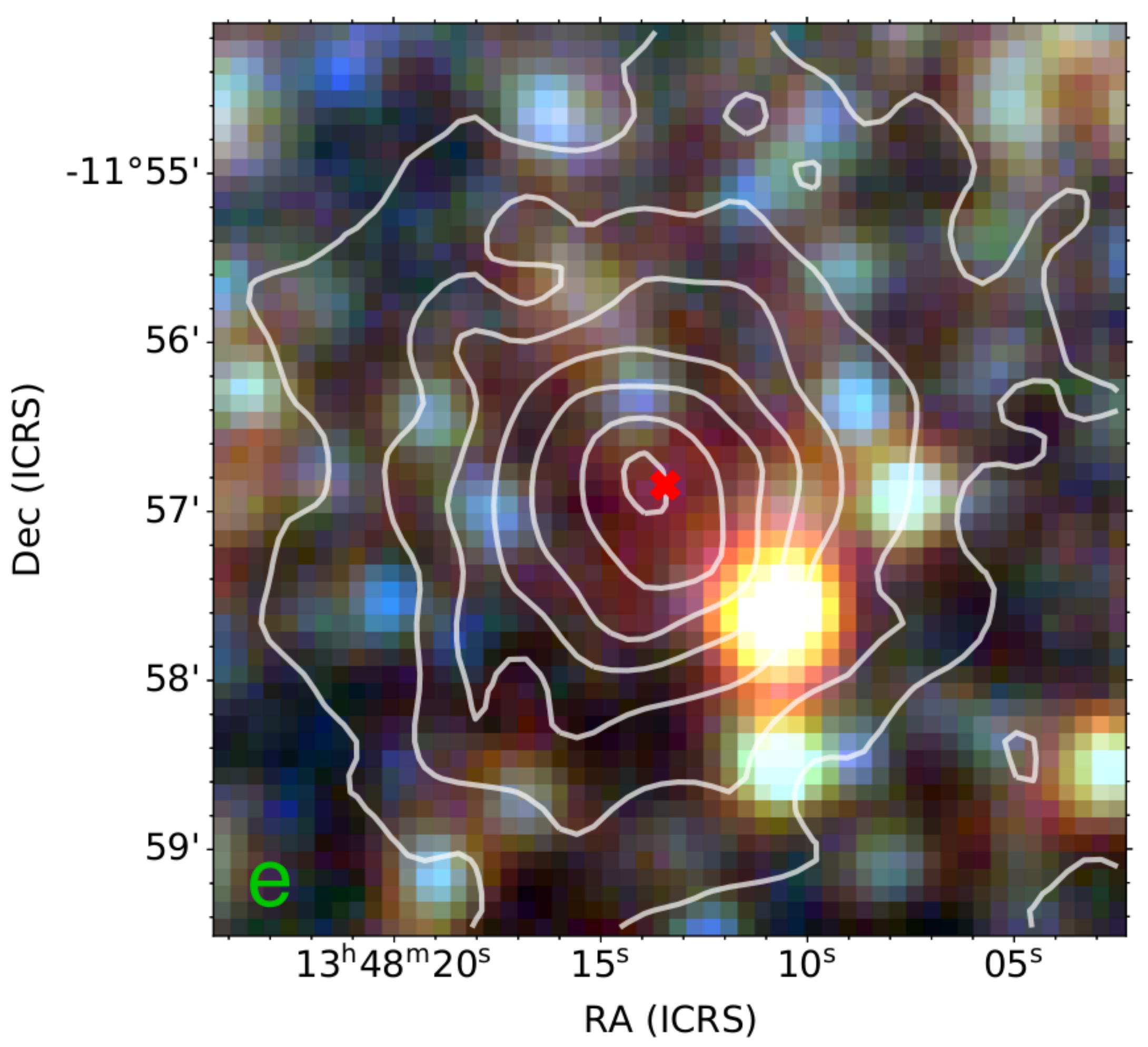}
        \caption{All images show the same 6x6 sq. arcmin field of view. a) SPIRE three-band false color image. The green box indicates where we fit for the SZ effect amplitude, is 4\arcmin\ $\times$ 4\arcmin\ in size, and is centered on the X-ray center. b) \textit{Chandra} X-ray surface brightness image with logarithmic scaling and smoothing applied. c) \textit{HST} CLASH optical image with lensing model in red/blue contours. d) Bolocam SZ effect image. The circle indicates $R_{2500}$, within which we compute the average SZ effect signal brightness. e) SPIRE three-band false color image showing a contour overlay of the \textit{Bolocam} image. The red cross denotes the X-ray center. Given the astrometric uncertainties, the peak of the SZ effect in the \textit{Bolocam} image is coincident with the X-ray center.}
    \label{fig:fig1}
    \end{figure}

    \subsection{SPIRE} \label{sec:spire}
        SPIRE was an imaging photometer consisting of three focal plane arrays with approximate band centers at 600, 850, and 1200~GHz, which will be identified as PLW, PMW and PSW respectively for the remainder of this manuscript \citet{Griffin2010}. The point spread functions (PSFs) are well approximated as Gaussians, with full-width at half maxima (FWHMs) of 36\arcsec, 25\arcsec, and 18\arcsec. \rxj\ was observed to a depth corresponding to instrument noise levels of $\lesssim 2$~mJy beam$^{-1}$ for all three bands as part of the \emph{Herschel} Multi-Tiered Extragalactic Survey \citep{Oliver10, Oliver12}. 
        
        The SPIRE science archive data used in this analysis were processed through the SMAP/SPIRE-HerMES Iterative Map Maker (SHIM) following the description in \citet{Levenson10}. SHIM is favored over the \textit{Herschel}-provided tools because it is optimized to separate large-scale correlated noise from the signal, making it better suited for the study of extended emission including the SZ effect. The transfer functions for diffuse astronomical signals were estimated using the methods described in \citet{Viero13}. The typical difference between the input and output maps due to high-pass filtering is $\lesssim 1$~per cent. This is sub-dominant to the $\sim 5$~per cent absolute calibration uncertainty for SPIRE (see Section~\ref{sec:error}), so we assume the map-space signal transfer function is unity (other than the overall mean signal level, which is not preserved by SHIM).
        
        
    \subsection{Bolocam} \label{sec:bolocam}
    
        We use Bolocam photometric imaging data collected at an SZ-emission weighted band center of 139~GHz with an approximately Gaussian PSF with a FWHM of 59.2\arcsec. The square images are 14\arcmin\ in size, have an overall calibration uncertainty of 1.7~per cent, and astrometry accurate to 5\arcsec. To remove atmospheric fluctuations from the data, a template subtraction and high-pass filter are applied to the data timestreams. This results in an effective angular high-pass filtering of the SZ effect signal, which has been accurately characterized. To obtain an estimate of the SZ effect surface brightness, we follow the general method described by \citet[and references therein]{Sayers2019}, which we briefly summarize here.
        
        First, an elliptical generalized NFW (gNFW) profile \citep{Nagai2007}, using the power-law exponents from \citet{Arnaud2010}, is fit to the combination of the Bolocam data and the {\it Planck} MILCA $y$-map \citep{Sayers2016, Planck2016_XXII}. As noted in \citet{Sayers2016}, the addition of the {\it Planck} data significantly improves the quality of this fit by constraining the SZ effect signal on large angular scales. The PSF of each instrument, along with the Bolocam filtering, are fully accounted for in these fits. Next, the fitted elliptical gNFW model is used to generate a 2-dimensional angular template of the SZ effect signal. This template is then fit to the Bolocam data, with its normalization as the only free parameter, to obtain the average surface brightness of the SZ effect signal within \rtwofive\ at a frequency of 139~GHz. \rtwofive\ corresponds to the spherical radius enclosing an average density 2500 times the critical density of the universe, and in \rxj\ is measured to be $0.71 \sim$ Mpc based on the analysis of \citet{Czakon2015}. We note that the technique used to obtain \rtwofive\ in \citet{Czakon2015} was based on a generalized scaling relation intended to be applicable to large, heterogeneous galaxy cluster samples. For highly relaxed objects, like \rxj, more accurate methods are available \citep[see, e.g.,][who find a value of 0.80~Mpc]{Mantz2014}. However, given that \rtwofive\ is employed in this work solely as a convenient aperture size that is well-matched to the observational data, and that \citet{Czakon2015} had previously computed aperture photometry values from Bolocam data within \rtwofive $\sim =0.71$ Mpc, we retain that value for this analysis.
        
        In performing the above fits, we assume the map-space noise in the Bolocam data can be described using a diagonal covariance matrix (i.e., there is no correlated noise between map pixels). Since this assumption is imperfect, we characterize the uncertainty on the SZ effect surface brightness by performing analogous fits to a set of 100 noise realizations. These noise realizations are identical to those used by \cite{Sayers2019}, with the addition of a kinematic SZ effect signal (see Section~\ref{sec:bolocam_astro}). We find that the distribution of SZ effect surface brightness values obtained from the fits to these noise realizations is approximately Gaussian, and so we assign an uncertainty to the measured SZ effect surface brightness in the observed data based on the standard deviation of this distribution.
    
    \subsection{Chandra} \label{sec:xray}
        ICM temperatures measured using the rSZ effect are approximately pressure-weighted \citep[e.g.,][]{Kay2008}, and so we compute an analogous temperature from the available {\it Chandra} X-ray spectroscopic imaging \footnote{For an example of the biases that can result from comparing the standard emission-weighted X-ray spectroscopic temperature to rSZ effect temperatures, see \citet{Lee2020}.}. The procedure for reducing and cleaning the data is described in \citet{Mantz2014,Mantz2015a}, although we use more recent versions of the {\it Chandra} analysis software and calibration files for this analysis (\texttt{CIAO} version 4.9 and \texttt{CALDB} version 4.7.4). From these data, we obtain deprojected density and temperature profiles using the techniques detailed in \citet{Mantz2014,Mantz2016}. Using these profiles, we then compute the pressure-weighted mean temperature of the ICM within a cylindrical volume defined by \rtwofive\ in the plane of the sky and the maximum radial extent probed by the X-ray data along the line of sight, corresponding to $\sim 1.8$~R$_{2500}$. As part of this calculation, we also apply the empirically-derived temperature calibration bias from \citet{Wan2021}, equal to $0.09 \pm 0.13$. With this calibration correction, we obtain a value of \Tpwxtwofive\ $= 17.4 \pm 2.3$~keV.
   
    \subsection{HST} \label{sec:hst}
    
        The galaxy cluster \rxj\ was extensively imaged with \emph{Hubble} as part of the Cluster Lensing and Supernova {\it HST} (CLASH) survey \citep{Postman2012,Koekemoer_2011}. The {\it HST} images were used to construct mass models according to the \texttt{PIEMDeNFW} formalism \citep{Zitrin_2015}. This parametric model incorporates elliptical NFW dark matter profiles to model the cluster dark matter halo, and double pseudo isothermal elliptical mass distributions to model cluster member galaxies. In order to construct a model that can be used over the full extent of the SPIRE maps, the best-fit parameters from the \texttt{PIEMDeNFW} model were used to extrapolate onto a larger grid \citep[for more details, see][]{Sayers2019}. These grids have sizes of 16\arcmin\ $\times$ 16\arcmin\ and a angular resolution of 0.25\arcsec. 

\section{Multi-Component Map Fitting}\label{sec:map_fit}

A significant challenge to measuring the SZ effect from observations in the sub-mm/mm is constraining a signal that is contaminated with emission from dusty star-forming galaxies (DSFGs), including background DSFGs that are lensed by the cluster, galactic cirrus dust emission and potentially other components such as diffuse dust emission from the cluster itself \citep{Vogelsberger19, Erler18,Planck_2016_ccde,Planck_2016_ccde2}. These spatially coincident signals can be disentangled through prior information on the spatial morphology and underlying spectral energy distributions (SEDs) of the emission components.

Spectral information has previously been used through multiband matched filtering to minimize the contribution from unwanted signals in a stacking analysis of galaxy clusters in \textit{Planck} maps \citep{Erler18}. However, in general {\it Planck} does not resolve the galaxy clusters, and so the possibility of using spatial information is not available. SPIRE data, however, permit detailed spatial-spectral modeling to separate the unresolved point-like DSFGs from the diffuse SZ effect and cirrus dust emission. The combination of angular resolution, instrument sensitivity and the DSFG luminosity function set intrinsic limits on the efficacy of this type of approach, but for certain observations the gains in robustness and constraining power can be significant. 

The multi-component fitting method employed in this work builds on the hierarchical modeling framework of probabilistic cataloging, or PCAT
\citep{Brewer12, Daylan17, Portillo17, Feder20}. PCAT is designed to explore the space of catalog models consistent with an observed image by fitting a Poisson mixture model directly to the data, where each component represents a point-like source. As a Bayesian hierarchical model, probabilistic cataloging combines prior information about the source population of interest with the data likelihood to estimate a posterior distribution of point-like source models consistent with our prior expectations and the data. This is represented as an ensemble of catalogs that naturally encode the often complicated model uncertainties that can arise in confusion-limited observations (such as from SPIRE) and crowded fields in general.

The major utility of PCAT as it applies to this analysis is that a point source model can be estimated directly from the data in such a way that marginalization over the point source model enables better estimation of a correlated signal of interest, where in this case the correlated signal is the SZ effect. This includes marginalization over uncertainties associated with point source positions and flux densities, as well as uncertainty due to our ignorance of the true number of DSFGs (down to a given flux density) in the observed field. This is necessary in the absence of ancillary data from deeper observations with finer angular resolution. In this section, we describe the forward model used in probabilistic cataloging and the extension to observations of the SZ effect toward galaxy clusters.

\subsection{Generative model}
\label{sec:gen_modl}

For an image with dimension $(\textrm{W}, \textrm{H})$ observed in band $b$, the surface brightness sampled by pixel $(i,j)$ is written as a sum over point-like and diffuse components convolved with the PSF $\mathcal{P}$:
\begin{equation}
\begin{split}
\lambda_{ij}^b = \mathcal{P}^b \circledast \Bigl[\sum_{n=1}^{\textrm{N}_{\textrm{src}}} S_{n}^b \delta(x_i-x_{n}^b,y_j-y_{n}^b) \; + \\ \textrm{A}_{b}^{\textrm{SZ}} I^{\textrm{SZ}}_b(x_i,y_j) \: + \: B_{ij}^b \Bigr].
\end{split}
\label{eq:gen_model}
\end{equation}
The first sum term denotes the contribution from \nsrc\ point-like sources with flux densities $\lbrace S_{n}\rbrace_{n=1}^{\textrm{N}_{\textrm{src}}}$ and positions $\lbrace x_n\rbrace_{n=1}^{\textrm{N}_{\textrm{src}}}$, and assumes that the galaxies in our image are well represented as point sources, which is reasonable given the SPIRE PSF size and the galaxy redshift distribution. The second term is the spatially extended SZ effect, included in our model as an angular template $I_{SZ}^b$ scaled by amplitude \ASZb, with the angular template constructed from the elliptical gNFW fit to Bolocam and {\it Planck} (see Section~\ref{sec:bolocam}).
The last term, $B_{ij}^b$, captures additional diffuse components in a non-parametric fashion through the addition of a two-dimensional truncated Fourier series:
\begin{equation}
    B_{ij}^b = B_{0}^{b} + \sum_{n_x=1}^{N_m}\sum_{n_y=1}^{N_m}\pmb{\beta}_{n_x n_y}\cdot \pmb{\mathcal{F}}_{ij}^{n_x n_y},
\end{equation}
where $\pmb{\mathcal{F}}_{ij}^{n_x n_y}$ is a vector of values from Fourier components with wavevector $(k_x, k_y) = (\textrm{W}/n_x,\textrm{H}/n_y)$ evaluated at pixel $(i,j)$:
\begin{equation}
    \pmb{\mathcal{F}}_{ij}^{n_x n_y} = \left(\begin{array}{c} \sin\left(\frac{n_x\pi x_j}{\textrm{W}}\right)\sin\left(\frac{n_y\pi y_j}{\textrm{H}}\right) \\ 
    \sin\left(\frac{n_x\pi x_j}{\textrm{W}}\right)\cos\left(\frac{n_y\pi y_j}{\textrm{H}}\right)
    \\
    \cos\left(\frac{n_x\pi x_j}{\textrm{W}}\right)\sin\left(\frac{n_y\pi y_j}{\textrm{H}}\right)
    \\
    \cos\left(\frac{n_x\pi x_j}{\textrm{W}}\right)\cos\left(\frac{n_y\pi y_j}{\textrm{H}}\right)
    \end{array}\right)
\end{equation}
and $\pmb{\beta}_{n_x n_y}$ are the component amplitudes. Including all four components in $\pmb{\mathcal{F}}^{n_x n_y}$ is necessary because we do not wish to impose boundary conditions on the diffuse model. We determine an appropriate truncation scale for the Fourier component model empirically by minimizing scatter on the inferred SZ effect template amplitudes recovered from mock observations (see Section~\ref{sec:mock_pipe}). In general, the choice of truncation scale depends on both the power spectrum of the diffuse signal contamination and the scale of the PSF, which acts as a low-pass filter through its convolution with $B_{ij}^b$. The mean surface brightness of each image, captured by \Bzero, is not physically meaningful because the SPIRE maps are not absolutely calibrated, and so \Bzero\ is treated as a nuisance parameter in our model.


\subsection{Data likelihood and priors}

Our data likelihood is assumed to be Gaussian and is written in map space as a product over pixels:
\begin{equation}
\mathcal{L} = \prod_{b=1}^{\textrm{B}}\prod_{i=1}^{\textrm{W}}\prod_{j=1}^{\textrm{H}} \frac{1}{\sqrt{2\pi\vec{\sigma}^2_b}}\exp\left(-\frac{(\vec{d}_b-\vec{\lambda}_b)^2}{2\vec{\sigma}^2_b}\right).
\label{likelihood}
\end{equation}
where $\vec{d}_b$ is the observed data vector and $\vec{\lambda}_b$ is the generated model. The noise model that sets the pixel-wise variance $\sigma_b^2$ in Eq. \eqref{likelihood} is estimated from the SPIRE timestream data following the procedure described in \citet{Viero13}. The above expression then reduces to
\begin{equation}
\log \mathcal{L} \approx \sum_{b=1}^{\textrm{B}}\sum_{i=1}^{\textrm{W}}\sum_{j=1}^{\textrm{H}} - \frac{(\vec{d}_b-\vec{\lambda}_b)^2}{2\vec{\sigma}_b^2}.
\label{eq:logL}
\end{equation}

We assume galaxies are randomly distributed on the sky, placing uniform priors on source positions, although we note this is an imperfect assumption mainly due to the spatial inhomogeneities resulting from gravitational lensing of CIB sources by the galaxy cluster. The multiband flux density prior $\pi$ is factorized into a flux density prior for the shortest wavelength band at 250$\mu$m (PSW) and color priors for the remaining two bands:
\begin{equation}
    \pi(\vec{S}) = \pi(S_{\textrm{PSW}})\pi\left(\frac{S_{\textrm{PMW}}}{S_{\textrm{PSW}}}\right)\pi\left(\frac{S_{\textrm{PLW}}}{S_{\textrm{PSW}}}\right).
\end{equation}
We assume the source flux density distribution follows a power law:
\begin{equation}
    \pi(S_{\textrm{PSW}}) \propto \left(\frac{S_{\textrm{PSW}}}{S^{\textrm{min}}_{\textrm{PSW}}}\right)^{-\alpha}
\end{equation}
and $\alpha = 3.0$ \citep{Casey14}. The color priors are modeled as Gaussian with means consistent with the typical DSFG SED. The widths of the color priors are optimized along with other hyperparameters using the mock observations described in Section \ref{sec:mock_pipe} to minimize scatter in the inferred SZ effect brightness. The minimum source flux density permitted by the model is determined in a similar fashion and is fixed to $S_{\textrm{PSW}}^{\rm min} = 5$ mJy.
The diffuse cirrus model is represented in image space as a linear combination of templates (one template per Fourier component), and the coefficients of those templates are sampled with uniform priors. Lastly, while the SZ effect increment between PMW and PLW has a well-defined range of colors for plausible temperatures, we choose to place independent, uniform priors on the SPIRE SZ effect template amplitudes to capture potential systematic effects that could bias the inferred surface brightness values.

By sampling the space of models consistent with observed data $D$ using Eq. \eqref{likelihood} and regularizing the set of solutions with suitable priors, we can compute the posterior distribution of astrophysical models, $P(\mathcal{M}|D)$, through Bayes' rule:
\begin{equation}
    P(\mathcal{M}|D) = \frac{\mathcal{L}(D|\mathcal{M})\pi(\mathcal{M})}{P(D)} \propto \mathcal{L}(D|\mathcal{M})\pi(\mathcal{M}).
\end{equation}
Here, $\mathcal{L}(D|\mathcal{M})$ is the likelihood of observing data $D$ given astrophysical signal model $\mathcal{M}$. PCAT uses a Markov Chain Monte Carlo (MCMC) sampler which has been optimized to efficiently explore the posterior distribution of catalogs consistent with image data. Details on implementation and the sampling algorithm can be found in \cite{Portillo17} and \cite{Feder20}. The extension of probabilistic cataloging that performs joint reconstruction with diffuse signal components, \texttt{PCAT-DE}, uses the same Metropolis-Hastings algorithm to sample mean signal level and template components in addition to a point source model.

\subsection{Fitting procedure}
\label{sec:fitting_procedure}

Our procedure to fit the model to multiband SPIRE image data happens in two steps.  First, \texttt{PCAT-DE} is run on the PSW-only data to determine a spatial model for cirrus dust emission. A sixth-order Fourier component model is fit to the data ($\theta_{\textrm{min}}\sim 1.6\arcmin$, $n_{param}=144$) along with the point source model and mean surface brightness level.  Second, \texttt{PCAT-DE} is run on all three bands simultaneously, with the shape of the Fourier component model fixed. The Fourier component model from PSW is then scaled to PMW and PLW, assuming a constant cirrus SED across the field of view and constant \Bzero. The assumed cirrus SED is taken from SPIRE observations of the H-ATLAS SDP field \citep{Bracco11}. In this step the SZ templates for PMW and PLW are added to and fit jointly with the rest of the model.

This two-step procedure is performed in order to mitigate the effect of degeneracies between the cirrus model and the other signal components. The spectrum of cirrus dust is well constrained and is brightest at short wavelengths, so determining the spatial structure of cirrus using PSW data alone is sufficient. The single-band fit also avoids possible bias of the cirrus dust model by the SZ effect itself, which is also spatially extended. The SZ effect template amplitudes are floated for PMW and PLW, but since the the SZ effect brightness is close to zero at high frequencies we fix \ASZpsw\ to the value predicated from the combination of the measured Bolocam SZ effect brightness \dIbolo\ and {\it Chandra} \Tpwxtwofive, with \dIpsw\ =0.005~MJy/sr. This eliminates the possibility of CIB and/or cirrus emission in the PSW map from being incorrectly modeled as SZ effect signal. 

Probabilistic cataloging is computationally expensive compared to other point source detection/extraction algorithms. This is because both the number of parameters in PCAT's forward model and the degrees of freedom in the image data are large. 
To ensure we recover a well-sampled posterior on $\lbrace$ \ASZb\ $\rbrace$, we restrict our multi-component map fitting procedure to the $4\arcmin \times 4\arcmin$ region centered on the SZ-defined cluster centroid. However, we compute the best fit mean level for each band using the larger $10\arcmin \times 10\arcmin$ maps and then fix those values in fits of the $4\arcmin \times 4\arcmin$ maps.
We confirm through tests on mock data that this procedure does not bias our SZ estimates, but we do include an additional statistical uncertainty determined by how much the inferred SZ template amplitudes vary using a range of mean surface brightness levels consistent with our $10\arcmin \times 10\arcmin$ map fits (see \S\ref{sec:error}). Crucially, fits on the larger maps are done using the full model from Eq. \eqref{eq:gen_model}, which allows for unbiased recovery of the mean level in each band. 

The outputs of this fitting process comprise: a catalog of three-band point source flux densities and associated errors, three-band Fourier component amplitudes of the cirrus emission, and SZ effect template amplitudes in PMW and PLW bands expressed as a fitted brightness amplitude \ASZb. These surface brightnesses can then be corrected for the relevant biases and combined with the measured Bolocam surface brightness to constrain the SZ effect parameters \ytwofive\ and \Ttwofive, as detailed in Section \ref{sS:SZlike}.

\section{Mock Observation Pipeline} \label{sec:mock_pipe}

\begin{figure*}
    \centering
    \includegraphics[width=0.95\linewidth]{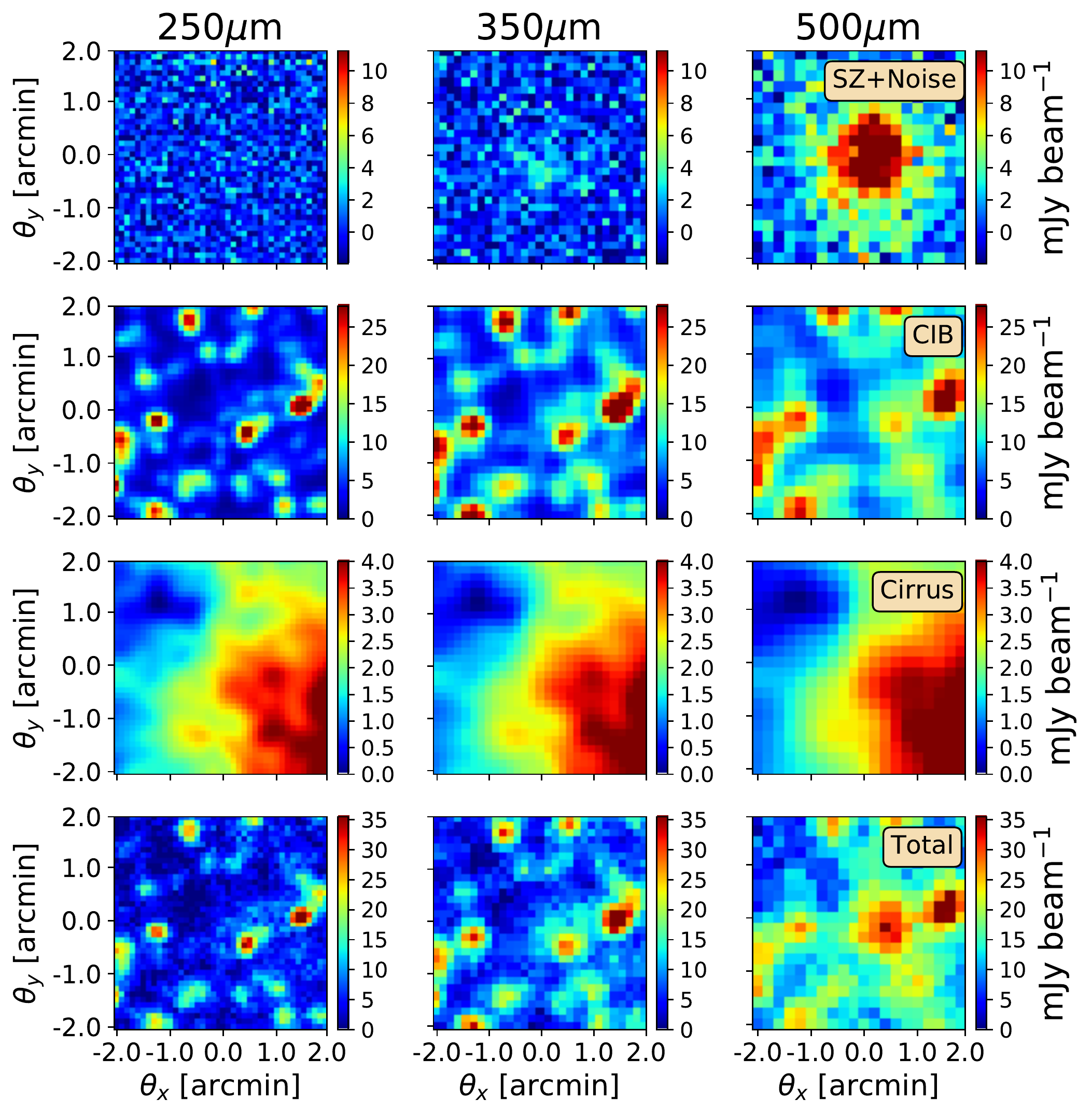}
    \caption{An example of one mock SPIRE observation of the galaxy cluster \rxj\ in PSW (left), PMW (middle) and PLW (right).}
    \label{fig:mock_realization}
\end{figure*}

In order to validate our analysis pipeline, assess biases on the inferred SZ effect signal, and quantify uncertainties associated with instrument noise and astrophysical contaminants, synthetic multiband SPIRE maps of \rxj\ are generated. The components in these maps include instrument noise, the SZ effect,  diffuse dust emission from galactic cirrus, and random realizations of the CIB that include the effects of gravitational lensing. In the following subsections we describe how these various components are generated. The individual and combined signal components for the three SPIRE bands are shown in Figure \ref{fig:mock_realization} for one mock observation.

\subsection{Instrument Noise}

We use the model described in \S\ref{sec:spire} to generate random instrument noise realizations for our mock observations. Specifically, we assume the noise fluctuations to be Gaussian and uncorrelated between map pixels, with per pixel uncertainties determined by the detectors' time stream variance and integration time in each pixel.  The spatial scan pattern of SPIRE results in a non-uniform integration time across the field of view that is largest in the central region and decreases towards the edges \citep{Oliver12}.

\subsection{Thermal SZ effect}

To include the SZ effect signal in our mock images, we use the elliptical gNFW model described in \S\ref{sec:bolocam} as a spatial template. The normalization of this template in the SPIRE bands is computed using \texttt{SZpack} based on the Bolocam measured surface brightness at 139~GHz and the {\it Chandra}-measured pressure-weighted temperature. The resulting images are then convolved with the appropriate PSF for each SPIRE band.

\subsection{Diffuse foregrounds}

Initial fits to observed data in \rxj\ with PCAT, which assumed a mean signal level and not a more general diffuse component, produced image residuals with spatial structure on large angular scales. Data from IRAS \citep{Miville2005} and {\it Planck} \citep{Planck2016_XLVIII,Planck_dust_2015} show emission with a consistent spatial and spectral signature, implying that the residual seen in our images is due primarily to thermal dust emission from galactic cirrus. However, the Planck beam size is very coarse compared to our map size, meaning only the largest modes are captured. Indeed, the \emph{Planck}-interpolated maps were not sufficiently accurate as spatial templates to model the diffuse emission in the SPIRE maps. As PCAT has a relatively low minimum source flux density threshold, residual diffuse emission can be misattributed to low-significance point sources. This was also observed using \emph{Planck}-interpolated templates, motivating the Fourier component approach described in Section \ref{sec:gen_modl}.

Random realizations of cirrus emission are created by drawing Gaussian random fields with angular power spectra consistent with the observed cirrus signal in \rxj. In particular, the power spectrum is assumed to scale as $P(k_{\theta}) \propto k_{\theta}^{-2.6}$ \citep{Bracco11}, with a normalization set by the measured amplitude of the large-scale power spectrum of the {\it Planck} dust template from \citet{Planck_dust_2015}, interpolated to the SPIRE footprint and extrapolated to SPIRE frequencies using the {\it Planck}-estimated parameters of a modified blackbody SED. As with the other astronomical signals included in our mock images, the resulting realizations are convolved with the PSF appropriate to each SPIRE band.

\subsection{CIB}
\label{sec:cibmock}
The CIB, which is due primarily to DSFGs but also active galactic nuclei (AGN), is the brightest astrophysical source of emission at high galactic latitudes \citep[][]{Hauser2001}. The depth of these observations and the angular resolution of SPIRE cause the image-space pixel fluctuations to be dominated by faint, undetected CIB sources rather than the instrument noise, commonly referred to as ``confusion noise" \citep[][]{Nguyen_2010}. Further, the vast majority of bright galaxies in the SPIRE images are not associated with the cluster, and are instead located behind it \citep[][]{Rawle2014,Rawle2016}. Since \rxj\ is an efficient gravitational lens \citep[][]{Bradac2004, Zitrin_2015}, most of the CIB sources in the SPIRE image have been deflected and magnified. As a result, it is necessary to consider not just the bright sources that can be detected individually, but also the undetected CIB sources, many of which have been lensed by the galaxy cluster.

We create mock observations of the CIB in the SPIRE bands using a two-step process wherein: bright sources individually detected by PCAT are used to produce constrained CIB realizations and random realizations of the population of faint undetected sources are generated from an empirical model of the CIB, including the effects of gravitational lensing on sources behind \rxj.  We describe these steps in detail below.

First, at flux densities above $2\sigma_{\textrm{conf}}$, corresponding to 11.6, 12.6, and 13.6~mJy in PSW, PMW, and PLW \citep{Nguyen_2010}, PCAT detects sources with a completeness of $\gtrsim 90$\% and a false detection rate of $\lesssim 10$\% (Feder et al. 2021, in prep.). We thus expect sources above these thresholds in the PCAT catalog to accurately describe the observed sky. Therefore, to create a single mock observation of the CIB, we extract the positions and flux densities of all catalog sources brighter than these thresholds in at least one SPIRE band. By populating the mocks with different realizations from PCAT's catalog ensemble, we ensure that the ensemble of mocks encodes the measured uncertainties from blind source extraction/photometry of the CIB.

Second, we generate random catalogs of the positions, SPIRE flux densities, and redshifts of a set of CIB galaxies using the empirical model from \citet{Bethermin2012}. All of the sources behind the galaxy cluster are gravitationally lensed according to the mass model derived in Section~\ref{sec:hst}. To retain the correct CIB population statistics, we then remove all of the sources brighter than $2\sigma_{\textrm{conf}}$ in at least one SPIRE band from the lensed source catalog. For a single mock, a full catalog of CIB sources is thus obtained from the combination of bright sources in the PCAT catalog and fainter sources remaining in the \citet{Bethermin2012} catalog after this removal. The SPIRE PSF is then used to generate a mock observation from this catalog.

\begin{figure}[h]
    \centering
    \includegraphics[width=0.95\linewidth]{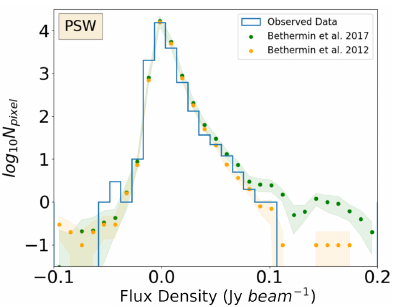}
    \label{fig:psw_hist}
    \includegraphics[width=0.95\linewidth]{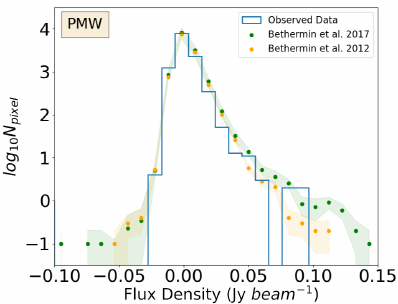}
    \label{fig:pmw_hist}
    \includegraphics[width=0.95\linewidth]{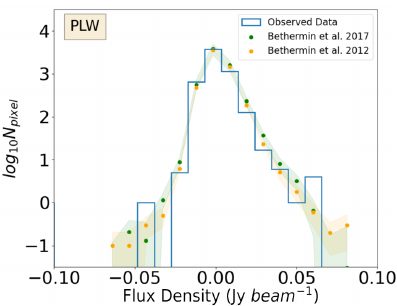}
    \caption{Pixel histograms of observed SPIRE maps and the mock observations created from both the \texttt{B12} and \texttt{B17} models. While the mock observations from \texttt{B12} (orange) show agreement with SPIRE observations, the \texttt{B17} model (green) produces an excess of bright pixels.}
    \label{fig:pix_hists}
\end{figure}

We note that the empirical model of \citet{Bethermin2012} was updated in \citet[][hereafter \texttt{B12} and \texttt{B17} respectively]{Bethermin_2017}, and we also created mock observations of the CIB using the more recent model. However, when the effects of gravitational lensing were included, we found a statistically significant excess of bright sources in mock observations created from the \texttt{B17} model compared to the observed data in the PSW and PMW bands (see Figure~\ref{fig:pix_hists}). Note that the \texttt{B12} and \texttt{B17} catalogs were lensed in an identical manner, and that the catalog generation process is independent of the lensing step. 

The underlying cause of this excess appears to be the gravitational lensing of numerous faint background sources by the galaxy cluster. However, it is unclear whether the excess is specific to the geometry of this particular gravitational lens, or if it generically appears towards massive galaxy clusters. Regardless of the specific cause of this excess when using the \texttt{B17} model, it is not observed in the mock observations created from the \texttt{B12} model, which show good agreement with the observed data. Therefore, we have used the older model for this analysis.

\section{Sources of Measurement Error}\label{sec:error}

In this section, the components of the mock observations described in Section \ref{sec:mock_pipe} are used to assess the impact of each type of signal (or noise) on the derived SZ results. We also isolate and study the effect of lensing on our inference by testing CIB realizations both with and without application of the lensing model.

    \subsection{Instrument noise}
            For Bolocam, we use the formalism described in \citet{Sayers2011} to produce random realizations that include both detector noise and fluctuations in the atmospheric emission. We note that the detector noise has an approximately flat spectrum, while the atmospheric fluctuations have a power spectrum that increases as a power law at low angular frequencies in the map.
        
            For SPIRE, the contribution from instrument noise is isolated by generating random realizations from the SPIRE noise model, adding a fiducial SZ effect signal to each realization and fitting the signal template to the data within our analysis pipeline. This may be interpreted as an estimate of the raw sensitivity of the measurement, in the absence of other systematics. Quantitative estimates of the uncertainties on the \dIb\ due to instrument noise are given in Table~\ref{tab:error}.
        
    \subsection{Astrophysical contamination - Bolocam} \label{sec:bolocam_astro}
            The Bolocam images contain a small, but non-negligible, amount of signal from unwanted astrophysical contaminants. As originally described in \citet{Sayers2011}, our noise model includes random realizations of the primary CMB anisotropies and CIB based on their measured angular power spectra. In addition, the signal from the bright AGN in the BCG is modeled and removed according to the procedures detailed in \citet{Sayers2013}. For this analysis, we also add a contribution to the noise model due to the kinematic SZ effect signal resulting from the (unknown) bulk line of sight velocity of the galaxy cluster. Following the convention of \citet{Mueller2015}, we assume a random velocity centered on zero with a standard deviation of 300~km~sec$^{-1}$ based on the simulations of \citet{Sheth2001}. In computing the signal from this velocity, we assume the ICM is isothermal with a temperature equal to the X-ray measured value of 17.4~keV. Quantitative estimates of the overall uncertainties on \dIbolo\ arising from these signals are given in Table~\ref{tab:error}.
        
    \subsection{Astrophysical contamination - SPIRE}
    
        The individual galaxies comprising the CIB contaminate our measurement of the SZ effect, particularly the spatially correlated emission arising due to gravitational lensing of the background population. Using the 100 mock CIB realizations described in Section \ref{sec:cibmock}, we estimate the associated uncertainty on the \dIb\ from the aggregate posterior for the values of \ASZb. Because the ensemble of CIB realizations is well approximated as a collection of independent, identically distributed draws from an underlying luminosity function, the aggregate posterior from these mocks should capture the effect of instrument noise, per-realization CIB model uncertainties, and any error due to intrinsic scatter from cosmic variance in the CIB. While the \texttt{B12} model does not include a clustering term, the fluctuations due to Poisson noise from the CIB dominate at the scales of interest to our map, and so is assumed to be negligible \citep[][see Figure 9]{Viero13}. 
        
        To quantify the contribution of cirrus dust contamination to our error budget, we compute the scatter on the inferred \dIb\ from an ensemble of 100 lensed mocks without cirrus, and then compare against the same mock observations with cirrus included. This constrains the effect of diffuse cirrus emission to the upper limits given in Table~\ref{tab:error}.
        
        Unlike in the Bolocam data, emission from the AGN in the BCG is relatively dim compared to the other signals in the SPIRE bands. Specifically, its brightness is approximately equal to the instrument noise per beam and almost two orders of magnitude dimmer than the integrated SZ effect signal within \rtwofive\ based on the SED fits from \citet{Komatsu1999} and \citet{Sayers2013}. Therefore, we have not attempted to specifically model the AGN emission in the SPIRE data.
        
        \subsubsection{Gravitational lensing}
        \label{sS:lensing}
    
        Gravitational lensing has a significant impact on our analysis of the CIB and the resulting SZ effect constraints. Uncertainties on the \dIb\ obtained from performing our analysis on lensed CIB mock realizations are presented in Table \ref{tab:error}, and equal to 0.014 and 0.012 MJy sr$^{-1}$ for PLW and PMW. In addition, we measure a significant bias in the value of the \dIb, equal to $-0.019$ and $-0.021$~MJy sr$^{-1}$ in the two bands. For comparison, we also performed our analysis using unlensed CIB mock realizations, obtaining slightly higher \dIb\ uncertainties of 0.016 and 0.014~MJy sr$^{-1}$ and significantly smaller (and positive) biases equal to $+0.007$~MJy sr$^{-1}$ in both bands.

        These differences between the lensed and unlensed mocks are due primarily to the effect of ``depletion" noted by \citet{Blain02}, where a lack of CIB emission is observed within the strong lensing region near the center of the galaxy cluster \citep[][]{Zemcov13,Sayers2019}. The effective subtraction of bright individual sources from the images using PCAT further enhances this depletion. Thus, the level of CIB emission coincident with the SZ effect signal is slightly lower than what is observed in unlensed regions of the sky. In addition, the inferred \dIb\ are biased low due to the on-average deficit of CIB emission coincident with the SZ effect signal. For our final analysis, we have corrected the values of \dIb\ for the measured bias, which, as noted above, is due primarily to lensing.
        
    \subsection{Instrumental flux calibration}
            The SPIRE data are first calibrated according to the procedure described by \citet{Bendo2013}. This calibration is then adjusted using the empirical cross-calibration factors between {\it Planck}-HFI and SPIRE determined by \citet{Bertincourt2016}. 
            As detailed by \citet{Bertincourt2016}, the {\it Planck}-HFI calibration has a statistical uncertainty of 1.1 and 1.4~per cent at 545 and 857~GHz, along with an absolute uncertainty of $\simeq 2$~per cent and $\simeq 5$~per cent, where the former is obtained from a measurement of the first two acoustic peaks in the primary CMB anisotropy power spectrum and the latter is obtained from the ESA2 planetary model of Uranus and the ESA3 planetary model of Neptune. In translating the {\it Planck}-HFI calibration to SPIRE, there is an additional $\simeq 4$~per cent uncertainty due to the SPIRE PSF calibration, along with a sub-per cent statistical uncertainty in the cross calibration. Adding these terms in quadrature, we estimate the absolute SPIRE calibration to be accurate to 4.6~per cent and 6.6~per cent for PLW and PMW. Furthermore, based on this calibration scheme, we expect a negligible correlation in the calibration uncertainty between PLW and PMW.
            
            The Bolocam data are calibrated according to the procedure described in \citet{Sayers2019}, which is accurate to 1.7~per cent. In brief, the empirical model derived in \cite{Sayers2012} is used to correct for variations in atomospheric transmission based on observing conditions. The planetary model of \citet{Griffin1993}, corrected based on the {\it Planck}-HFI planetary brightness measurements of \citet{Planck2017_LII}, is then employed to determine the absolute calibration.
            
    \subsection{Additional instrumental calibrations}
            
            While there are additional sources of potential systematic errors related to instrumental calibration (e.g., the measured spectral bandpasses, PSF shape measurements, astrometric corrections), all of these are likely to be sub-dominant to the flux calibration accuracy. In addition, many of these potential sources of systematic error have already been subsumed into the flux calibration model, and are thus largely accounted for. We therefore do not explicitly include them in our overall error budget.
        
        \begin{deluxetable*}{ccc}
            \tablenum{1}
            \tablecaption{Sources of Measurement Error\label{tab:error}}
            \tablewidth{0pt}
            \tablehead{
                \colhead{Source} & \colhead{\phantom{$<$~} Uncertainty on \dIb\ \phantom{$^{\textrm{a}}$}}  & \colhead{Notes} \\
                \colhead{} & \colhead{(MJy sr$^{-1}$)} & \colhead{}}
            \startdata
                \cutinhead{Bolocam}
                Instrument+Atmosphere & \phantom{$<$~}0.008\phantom{$^{\textrm{a}}$} & See \cite{Sayers2011} \\
                CMB+CIB & \phantom{$<$~}0.003\phantom{$^{\textrm{a}}$} & See \citet{Sayers2011} \\
                Kinematic SZE & \phantom{$<$~}0.007\phantom{$^{\textrm{a}}$} & Assuming $\pm 300$~km sec$^{-1}$ \citep{Sheth2001} \\
                Absolute Calibration & \phantom{$<$~}0.003\phantom{$^{\textrm{a}}$} & $\pm 1.7$ per cent, see \citet{Sayers2019} \\
                Total & \phantom{$<$~}0.012\phantom{$^{\textrm{a}}$} & Measured \dIbolo\ $ = -0.153$ MJy sr$^{-1}$ \\
                \cutinhead{SPIRE PLW} 
                Instrument & \phantom{$<$~}0.003\phantom{$^{\textrm{a}}$} & Mock SZ + Instrument Noise run through pipeline \\
                CIB & \phantom{$<$~}0.014$^{\textrm{a}}$ & Mean bias of $-0.019$ MJy sr$^{-1}$ (primarily due to lensing$^{\textrm{b}}$) \\
                Cirrus & \phantom{$<$}0.003 & Mocks with and without cirrus have consistent posteriors\\
                Absolute Calibration & \phantom{$<$~}0.003\phantom{$^{\textrm{a}}$} & $\pm 4.6$ per cent based on \citet{Bertincourt2016} \\
                Model fitting & \phantom{$<$~}0.002\phantom{$^{\textrm{a}}$} & Fixed mean signal level\\
                Total & \phantom{$<$~}0.014\phantom{$^{\textrm{a}}$} & Measured \dIplw\ $=0.104$ MJy sr$^{-1}$\\
                \cutinhead{SPIRE PMW} 
                Instrument & \phantom{$<$~}0.004\phantom{$^{\textrm{a}}$} & Mock SZ + Instrument Noise run through pipeline \\
                CIB & \phantom{$<$~}0.012$^{\textrm{a}}$  & Mean bias of $-0.021$ MJy sr$^{-1}$ (primarily due to lensing$^{\textrm{b}}$)\\
                Cirrus & \phantom{$<$}0.003 & Mocks with and without cirrus have consistent posteriors\\
                Absolute Calibration & \phantom{$<$~}0.002\phantom{$^{\textrm{a}}$} & $\pm 6.6$ per cent based on \citet{Bertincourt2016} \\
                Model fitting & \phantom{$<$~}0.003\phantom{$^{\textrm{a}}$} & Fixed mean signal level \\
                Total & \phantom{$<$~}0.013\phantom{$^{\textrm{a}}$} & Measured \dIpmw\ $=0.037$ MJy sr$^{-1}$ \\
            \enddata
            \tablecomments{Estimated uncertainty on the value of \dIb\ due to both instrumental and astrophysical sources (see text for more details). The uncertainties due to CIB fluctuations in the SPIRE PLW and PMW bands are highly correlated, see Figure~\ref{fig:corner}. \\ $^{\textrm{a}}$While the CIB-related uncertainty on \dIb\ is dominated by cosmic variance, a small amount is also related to fluctuations in the value of \nsrc\ in PCAT (corresponding to 0.004 MJy sr$^{-1}$ for PLW and 0.003 MJy sr$^{-1}$ for PMW). \\
            $^{\textrm{b}}$If the effects of gravitational lensing are not included in the mocks, then the uncertainty on \dIb\ attributed to variations in the CIB is slightly larger (0.016 MJy sr$^{-1}$ for PLW and 0.014 MJy sr$^{-1}$ for PMW). In addition, the magnitude of the mean bias on \dIb\ is significantly reduced and its value is slightly positive ($+0.007$ MJy sr$^{-1}$ for both PLW and PMW). Thus, the on-average decrease in CIB brightness near the galaxy cluster center due to gravitational lensing strongly biases the inferred \dIb\ to lower values while also minimally decreasing the associated uncertainty.}
        \end{deluxetable*}

\vspace{-1.0cm}
\section{Results}\label{sec:spec_fit}

\subsection{PCAT SPIRE results}

The quantities \dIpmw\ and \dIplw\ are inferred using a combination of the observed SPIRE data and the set of constrained mock observations. To ensure full coverage over the SZ posterior, we run 100 randomly initialized Markov chains on the observed data in parallel. Each chain is run for $5\times 10^6$ samples, which are then thinned by a factor of 1000. The \texttt{PCAT-DE} model provides a reasonably good but imperfect description of the data, with typical reduced-$\chi^2$ values, computed using the pixel-wise log-likelihood in Eq. \eqref{eq:logL} and the number of parameters at each step of the chain, in the range of 1.2--1.3. 

Running several independent chains allows us to assess the level of convergence in our model. We check that our chains have completed the ``burn-in" phase by visually inspecting the ensemble of trace plots and confirming that they are well mixed. We discard the first 50\% of samples and combine the remaining ones from all chains to produce the aggregate posterior distribution of \ASZb\ values. The mean auto-correlation lengths from the collection of observed chains are $\langle\tau\rangle_{\textrm{PMW}} = 6$  (thinned) samples and $\langle\tau\rangle_{\textrm{PLW}} = 22$,  which results in effective sample sizes large enough that Monte Carlo error on the results are $<1\%$. We further validate that our chains are well mixed using the Gelman-Rubin statistic, also known as the potential scale reduction factor (PSRF) \citep{GELMAN_RUBIN}. For the 100 randomly initialized chains run on the observed data, $\hat{R}=1.14$ for PMW and $\hat{R}=1.2$ for PLW, which suggests sufficient convergence to use the aggregated samples from all observed chains to compute an unbiased estimate of the posterior variance. By running several chains on individual mock observations, we measure $\hat{R} < 1.1$ and so we do not include any additional correction to the 100 chains run on the mocks, where only one chain is assigned to each mock realization.

\begin{figure}
    \centering
    \includegraphics[width=\linewidth]{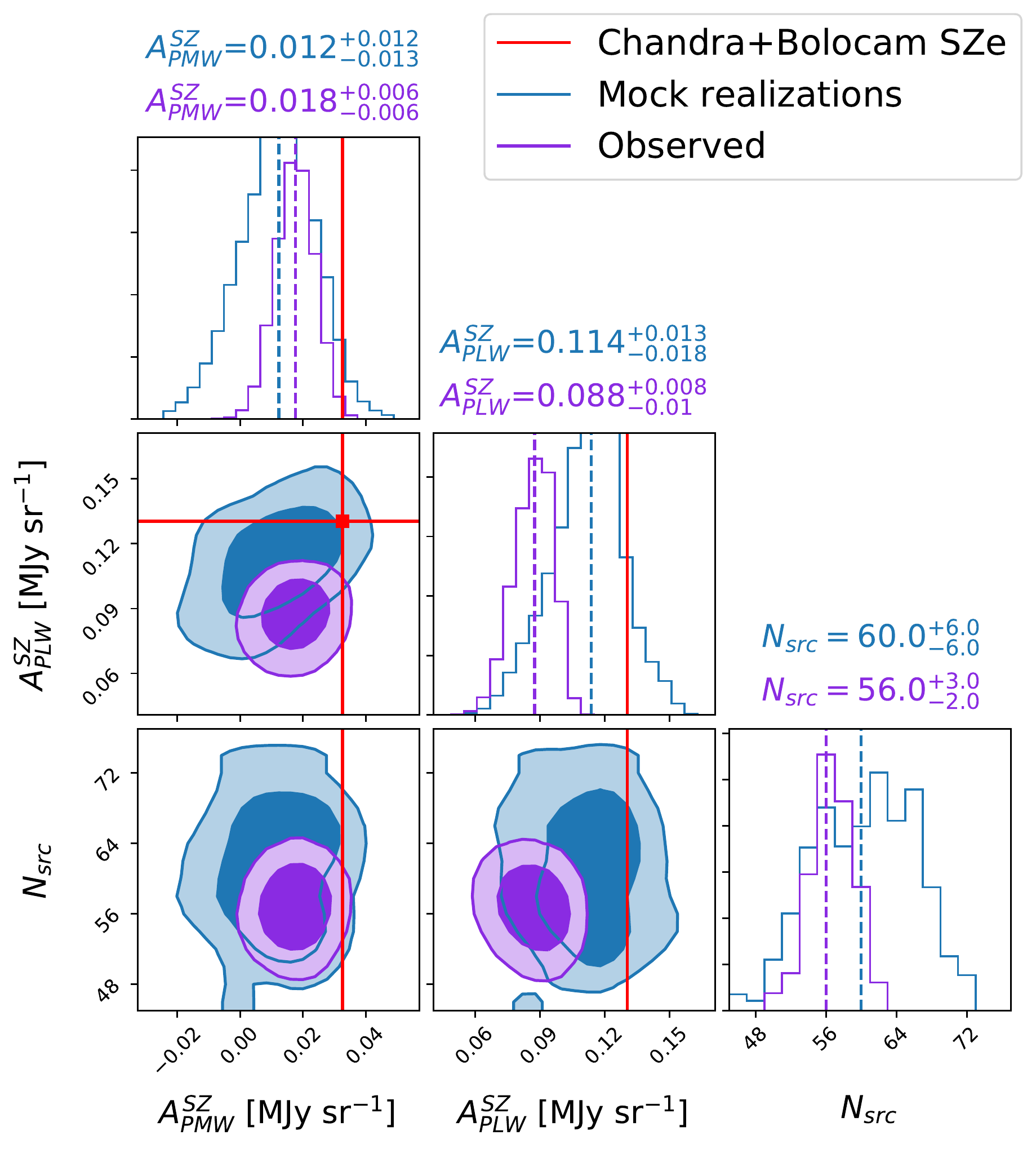}
    \caption{Marginalized posterior from 100 PCAT chains run on SPIRE maps of \rxj\ for SZ effect template amplitudes \ASZplw\ and \ASZpmw, and the inferred number of CIB sources \nsrc. Template amplitudes represent the average surface brightnesses within $R_{2500}$ and have units of MJy sr$^{-1}$. The SZ effect posteriors presented here are uncorrected for lensing depletion bias. The red crosshairs indicate the level of injected SZ effect signal used as ``ground truth" in our mock simulations to estimate the lensing depletion bias, which is taken from the measured \emph{Chandra} temperature of \Tpwxtwofive\ $=17.4$ keV combined with the measured Bolocam \dIbolo. Solid contours cover the 68 and 95~per cent credible regions of each posterior distribution.}
    \label{fig:corner}
\end{figure}

Figure \ref{fig:corner} shows the marginalized posterior on \ASZpmw\ and \ASZplw, along with the inferred number of point-like sources. Both the mock and observed posteriors on \ASZb\ in Fig.~\ref{fig:corner} are uncorrected for the bias due primarily to lensing depletion (see Section~\ref{sS:lensing}). In both the observed data and mock realizations, \ASZpmw\ and \ASZplw\ are positively correlated, with Pearson correlation coefficients of +0.18 and +0.52 respectively.
Since the SZ effect is independently observed across different SPIRE bands, the correlation between the two is due to the presence of correlations in the unresolved source population that are degenerate with the SZ effect signal. In the observed data, \ASZpmw\ and \nsrc\ appear uncorrelated. However, \ASZplw\ is anti-correlated with the number of detected CIB sources with a correlation coefficient of $-0.25$. This comports with the coarser PLW angular resolution, which results in the CIB being more spatially degenerate with extended emission from the SZ effect. The anti-correlation between \ASZplw\ and \nsrc\ is washed out in the aggregate mock posterior, where intrinsic scatter from cosmic variance in the CIB is dominant. 

Probing the cross-model covariance between the SZ effect and the union of point source models with varying \nsrc\ is straightforward in the framework of probabilistic cataloging, since for each SZ posterior sample there is an associated catalog of CIB sources. To assess the impact of our imperfect knowledge of \nsrc, we take the quadrature difference of the fully marginalized uncertainties on \ASZb\ with those from the conditional uncertainty assuming the median inferred \nsrc, i.e. $\sqrt{\sigma^2(\textrm{A}_b^{SZ})-\sigma^2(\textrm{A}_b^{SZ}|\textrm{med}(\textrm{N}_{\textrm{src}}))}$. By this metric, our results suggest that imperfect knowledge of \nsrc\ results in an uncertainty on \dIb\ that is $\sim 4$ times smaller than the uncertainty due to cosmic variance in the CIB.

Additional uncertainties due to the use of a fixed mean level estimated from the larger $10\arcmin \times 10\arcmin$ maps, inflating the constraints by 
and the convergence of the Markov chains are added as a random Gaussian component to the set of corrected samples, inflating the constraints on \dIpmw\ and \dIplw\ by 17\% and 20\%, respectively. After correcting for the lensing depletion bias, we measure the SZ effect brightness to be \dIplw\ $= 0.104 \pm 0.014$~MJy/sr and \dIpmw\ $= 0.037\pm 0.013$~MJy/sr, corresponding to 7.4$\sigma$ and 2.9$\sigma$ detections of the SZ effect, respectively.

\begin{figure}
    \centering
    \includegraphics[width=\linewidth]{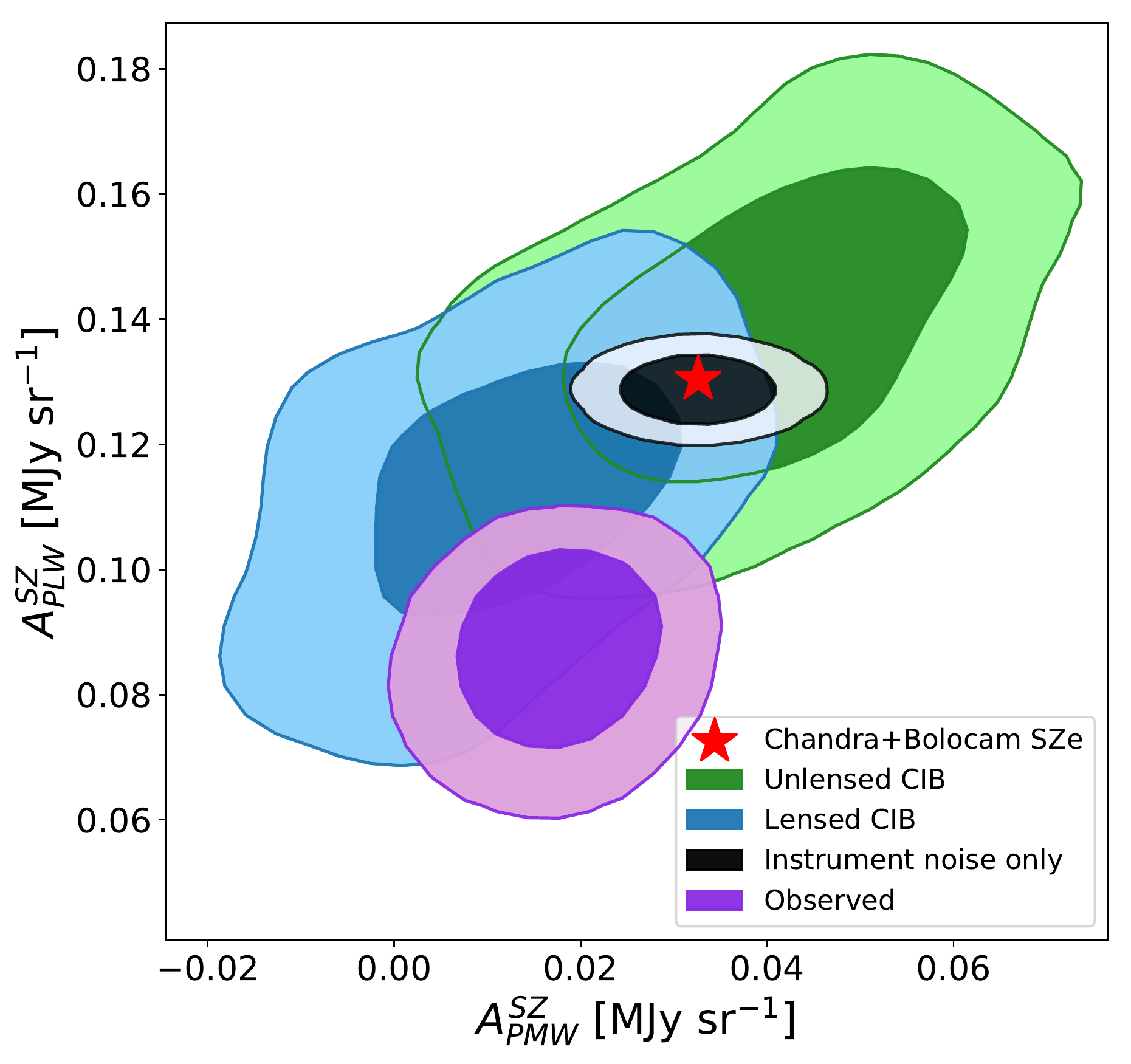}
    \caption{68 and 95~per cent credible regions for \ASZpmw\ and \ASZplw. The posteriors are composed from the distribution of template amplitudes sampled by \texttt{PCAT-DE}. Credible regions are shown for the observed \rxj\ SPIRE data (purple) and mock SPIRE maps with instrument noise only (black), instrument noise + unlensed CIB (green) and instrument noise + lensed CIB (blue). The red star indicates the injected SZ effect surface brightness values in the mock observations.}
    \label{fig:corner_compare}
\end{figure}

Finally, the difference between the posterior SZ effect samples and the \emph{Chandra}+Bolocam SZ effect amplitude injected into the mocks defines a bias distribution, which is used to propagate estimates from the observed (biased) value of \ASZb\ to the underlying \dIb\ (see Figure~\ref{fig:corner_compare} and also the discussion in Section~\ref{sS:lensing}). To test for model dependence in the bias, mock realizations with different injected SZ effect amplitudes were analyzed. For amplitudes spanning the range $[0,2]$ times the nominal \emph{Chandra}+Bolocam value, the mean bias is found to be constant within our measurement precision. Thus, it is valid to apply the linear bias correction to any measured values of \ASZb\ within this range. 

\subsection{SZ spectral likelihood analysis}\label{sS:SZlike}

    To obtain constraints on the average Comptonization parameter \ytwofive\ and ICM temperature \Ttwofive, we perform a likelihood analysis of our observed \dIb\ over a range of \ytwofive\ and \Ttwofive\ values, with \Ttwofive\ varying between $[0,60]$~keV. 
    The model \dIb\ values for each \ytwofive/\Ttwofive\ grid pair are computed using \texttt{SZpack} \citep{Chluba2012,Chluba2013}. In all cases, the effective band center of each \dIb\ used in this calculation was obtained by weighting the measured spectral bandpass of the instrument by the temperature-dependent shape of the SZ effect spectrum at the given value of \Ttwofive. The likelihood assumes independence between the Bolocam and SPIRE measurements of \dIb, which is a good approximation given that the CIB and cirrus signals are negligible in the Bolocam image and the primary CMB anisotropies and kinematic SZ effect signal are negligible in the SPIRE images.

    Our results using \texttt{PCAT-DE} demonstrate that the inferred distribution of \dIplw\ and \dIpmw\ values is non-Gaussian to such a degree that approximating the PDF with a Gaussian covariance matrix significantly distorts the final constraints. Instead, when computing the posterior $p(y,T|d\textrm{I})$ with Bayes' rule:
    \begin{equation}
    p(y,T|d\textrm{I}) \propto p(d\textrm{I}|y,T)p(y,T),
    \end{equation}
    the likelihood $p(d\textrm{I}|y,T)$ is evaluated directly from the gridded, bias-corrected samples of \dIplw\ and \dIpmw\ (also referred to as the ``empirical PDF"). The prior $p(y,T)$ places constraints on the range of spectra that are consistent with our underlying model. In our current analysis, no prior is placed on the amplitudes of the SZ templates, nor on their color, at the map-fitting level. While this helps in identifying and correcting for systematic effects, it does mean a subset of samples from the posterior fall outside the range of \dIb\ computed by \texttt{SZPack} for $0 \le $~\Ttwofive~$\le 60$ keV. These samples, which reside on the tails of the PDF, comprise only $\sim7\%$ of the full sample and are given zero weight in the final constraints.
    
\begin{figure}[t]
        \centering
        \includegraphics[width=\linewidth]{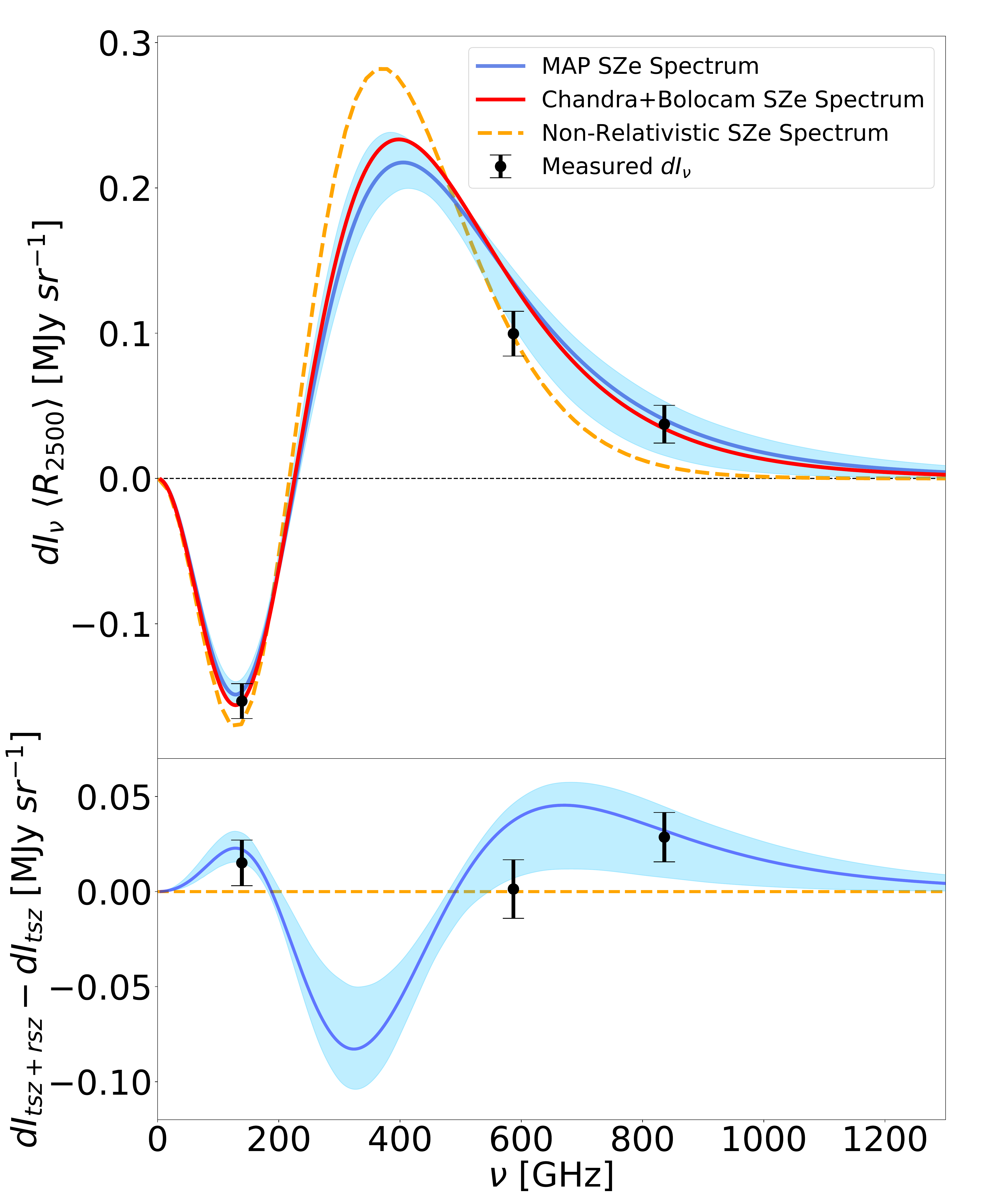}
        \caption{Top: SZ effect spectrum for the Bolocam+SPIRE MAP \ytwofive\ and \Ttwofive\ values (thin blue), along with the spectrum obtained from the Bolocam \dIbolo\ and {\it Chandra} \Tpwxtwofive\ (red). Also shown is the best-fit non-relativistic SZ effect spectrum (dashed orange). The blue shaded region bounds the range of SZ effect spectra within the 68~per cent credible region of \ytwofive\ and \Ttwofive. The \dIb\ values are shown as black points, with reported errors. Bottom: Residual spectra after subtraction of the non-relativistic SZ effect spectrum. Note that, due to the negative sign of the SZ effect in the Bolocam band, a measured \dIbolo\ below the MAP spectrum indicates a larger rather than a smaller SZ effect amplitude. The PSW band center is at approximately 1200~GHz, where there is very little SZ effect signal.}
        \label{fig:best_spec}
    \end{figure}

\begin{figure*}[t]
        \centering
        \includegraphics[width=\linewidth]{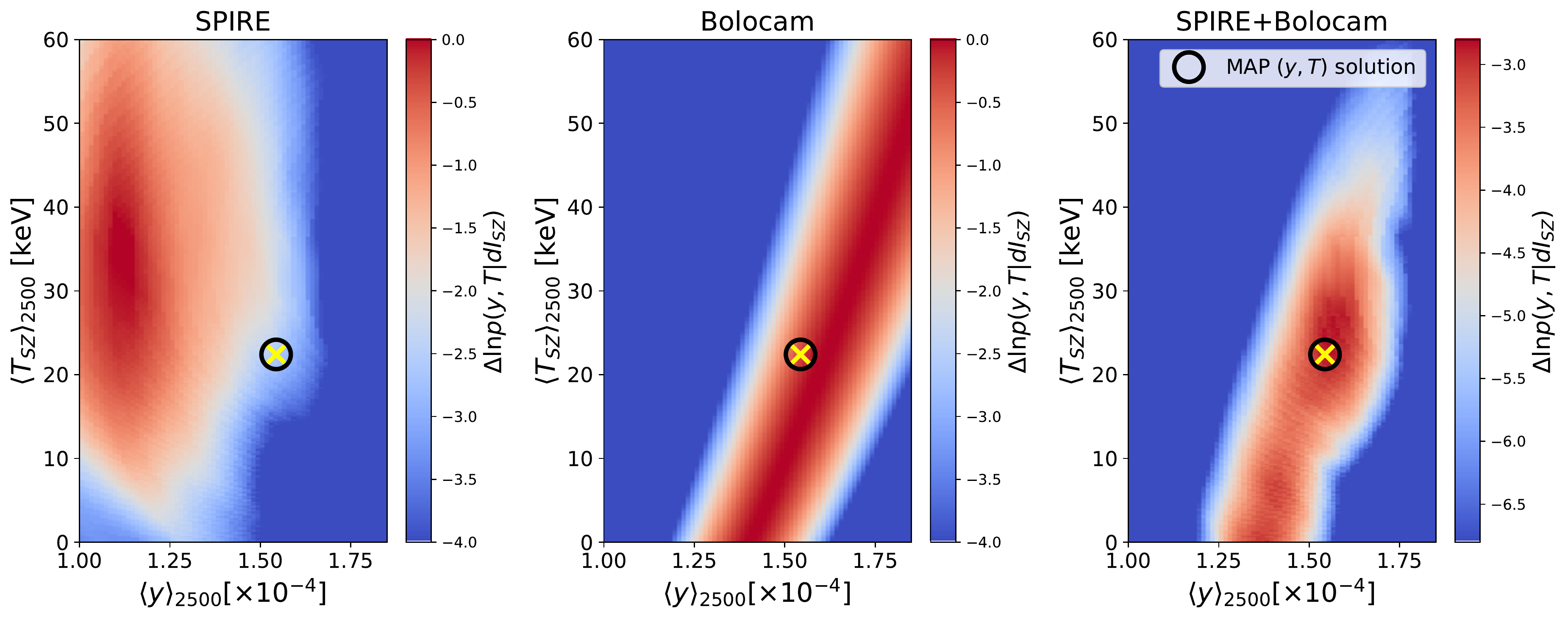}
        \caption{The joint posterior distribution (right) is computed as the product of the SPIRE posterior (left), which is sampled directly with \texttt{PCAT-DE}, and the Bolocam likelihood (middle) which is assumed to be Gausisan-distributed with respect to both parameters. The black circle + yellow cross indicates the position of the maximum \emph{a posteriori} (MAP) estimate of \ytwofive\ and \Ttwofive. Log-probabilities are plotted for visualization purposes.}
        \label{fig:joint_posterior}
    \end{figure*}

    The measured \dIb\ and constrained set of SZ effect spectra are shown in Figure \ref{fig:best_spec}, and the posterior on \ytwofive\ and \Ttwofive\ is shown in Figure \ref{fig:joint_posterior}. Looking separately at the contributions from Bolocam and SPIRE, one can see that the individual measurements suggest different constraints on \ytwofive\ and \Ttwofive. The Bolocam data constrain the parameters along a positively correlated axis in the \ytwofive/\Ttwofive\ plane. 
    The \ytwofive/\Ttwofive\ constraints using the empirical SZ posterior from SPIRE data have a more complicated structure. The preference of a smaller value of \ytwofive\ compared to Bolocam is largely driven by the value of \dIplw, which is approximately $\sim 2\sigma$ lower than the value expected from the combination of Bolocam and {\it Chandra}. While the Bolocam data are in good agreement with the maximum \emph{a posteriori} model estimate, the delta log-likelihood from the SPIRE points is $\Delta \ln\mathcal{L} = -2.4$, indicating tension with the spectral model.

    The joint posterior contains the most probability mass near the tails of the SPIRE-only posterior, and when combined with Bolocam data leads to bimodal constraints. This is due to the positive covariance between \dIplw\ and \dIpmw\ combined with the measured values falling on opposite sides of the SZ effect spectrum for that range of \ytwofive/\Ttwofive.  That is, the measured values scatter in an anti-correlated manner relative to the model while the covariance matrix implies a positive correlation, which leads to two local peaks in $p(d\textrm{I}|y,T)$. There is mild preference for the mode with larger values of the pair.   
    
    \begin{figure}
        \centering
        \includegraphics[width=\linewidth]{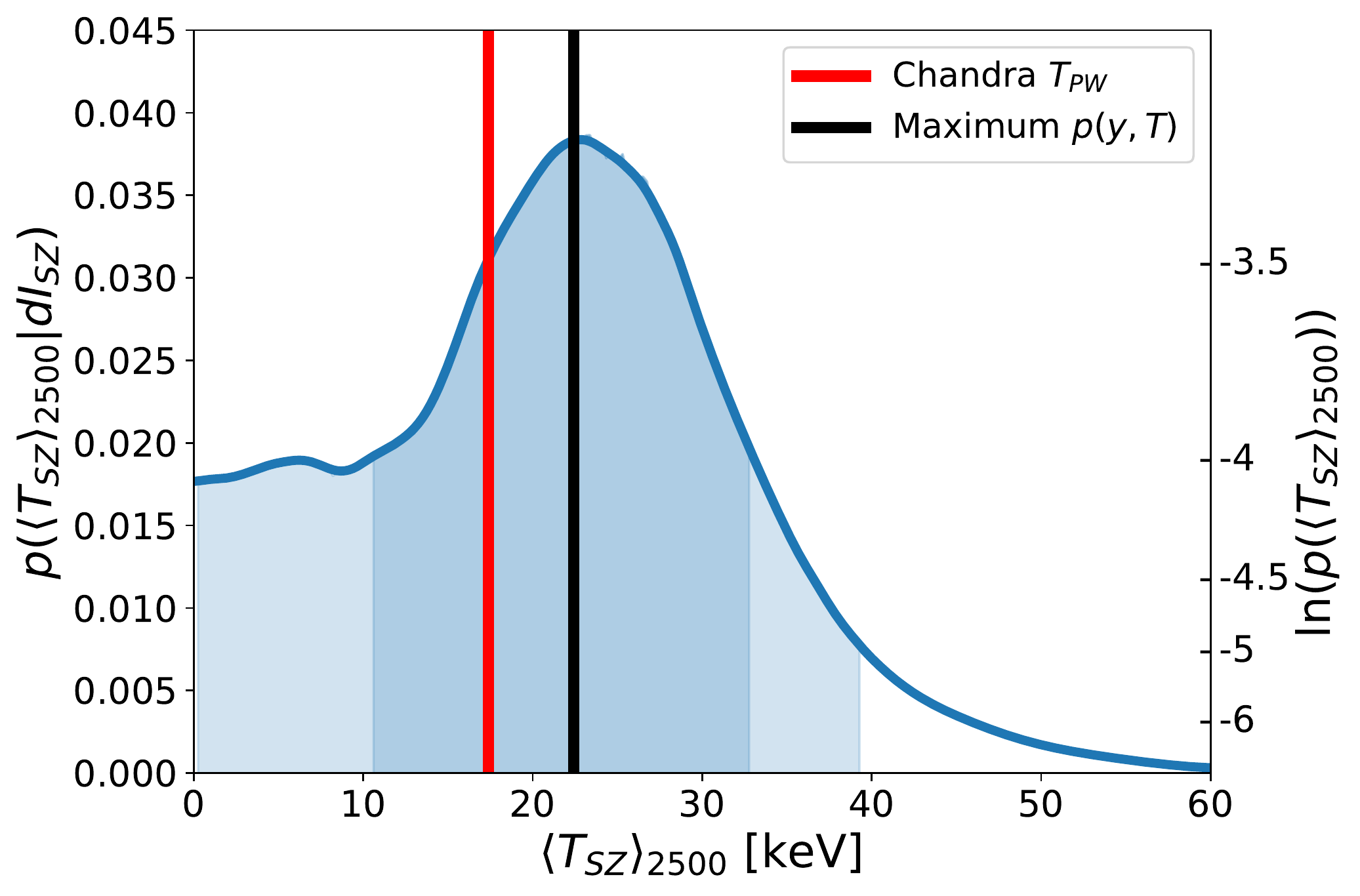}
        \caption{1D marginalized probability distribution for cluster temperature \Ttwofive, with joint MAP temperature (black) and credible intervals shaded in blue.}
        \label{fig:margin_t}
    \end{figure}
    
     \begin{deluxetable*}{cccc}
        \tablenum{2}
        \tablecaption{Summary of Results}
        \tablewidth{0pt}
        \tablehead{
            \colhead{Parameter} & \colhead{MAP value}  & \colhead{68 per cent credible interval} & \colhead{95 per cent credible interval}}
        \startdata
        \ytwofive\ $[\times 10^{-4}]$ & 1.56 & $1.42 <$ \ytwofive\ $< 1.63$ & $1.29 <$ \ytwofive\ $< 1.71$ \\
        \Ttwofive\ [keV] & 22.4 & $10.4 <$ \Ttwofive\ $< 33.0$ & $0.0 <$ \Ttwofive\ $< 39.5$
        \enddata
        \tablecomments{Fitted values of \ytwofive\ and \Ttwofive\ for \rxj.}
        \label{tab:MLE_table}
    \end{deluxetable*}
    \vspace{-0.8cm}
    Our maximum \textit{a posteriori} (MAP) estimate for the ICM temperature is \Ttwofive $= 22.4$ keV, with a 68\% highest posterior density credible interval (HDPI) of $10.4 <$ \Ttwofive\ $< 33.0$ keV (see Figure~\ref{fig:margin_t}). This estimate is consistent with the X-ray predicted, pressure-weighted temperature of \Tpwxtwofive\ $= 17.4 \pm 2.3$ keV, indicating good agreement between measurements. 
    
    To understand the sensitivity of our data in the absence of the bimodality noted above, we also repeat the same analysis using the \dIplw\ and \dIpmw\ values determined from the combination of Bolocam and {\it Chandra} (i.e., \dIplw\ $= 0.134$ MJy sr$^{-1}$ and \dIpmw\ $= 0.033$ MJy sr$^{-1}$). This eliminates the issue of positively-covariant values falling on opposite sides of the SZ effect spectrum, and removes the bimodality, while fully preserving the noise as characterized by the empirical PDF. Under these conditions, the posterior is single-peaked and the constraints marginalized over \ytwofive\ yield a MAP \Ttwofive\ $= 12.3$ keV with a 68~per cent credible interval of $5.8 <$ \Ttwofive\ $< 20.5$ keV. This represents a reduction in the range of the credible interval by 35~per cent compared to the observed data.
    
    If we instead repeat the analysis using the observed values of \dIb, but with randomized samples to remove the correlation between \dIplw\ and \dIpmw\ in the empirical PDF, then we find a 68~per cent credible interval of $5.1 <$ \Ttwofive\ $< 20.2$ keV. This is nearly identical to the interval obtained using \emph{Chandra}+Bolocam predicted \dIb\ values to eliminate the bimodality, and suggests that the \dIplw/\dIpmw\ correlation similarly degrades our constraints on \Ttwofive.

    As illustrated in Table~\ref{tab:error}, CIB fluctuations dominate the uncertainties on \dIplw\ and \dIpmw. To quantitatively assess the impact of the CIB on our derived \Ttwofive\ constraints, we also perform our analysis on the SZ + instrument noise only mock posterior (see Figure \ref{fig:corner_compare}), finding a MAP estimate of \Ttwofive=17.1 keV with a 68~per cent credible interval of $12.8 <$ \Ttwofive\ $< 19.0$ keV.

\vspace{0.5cm}
\section{Discussion}\label{sec:results}
    
    We have combined observations from SPIRE, Bolocam, and {\it Planck} to constrain the thermodynamic properties of the ICM in the galaxy cluster \rxj\ through a measurement of the rSZ effect. In order to probe the desired SZ effect signal, we accounted for significant contamination from unwanted astrophysical components, in particular the CIB. Not only is the CIB brighter than the SZ effect signal in the SPIRE bands, but the individual sources that comprise it have a range of SEDs. In addition, most CIB sources reside behind \rxj\ and have thus been deflected and magnified due to gravitational lensing. To properly characterize the effects of the CIB and other relevant signals such as galactic cirrus emission, we developed a mock observation pipeline to fully assess the uncertainties and biases in measuring the SZ effect brightness \dIb. Our mock pipeline includes a mass model derived from {\it HST}, allowing us to accurately capture the effects of gravitational lensing, which we found to be the largest source of bias on the derived values of \dIb.
    
    Rather than treating each astrophysical source in isolation, we employ an extension of the forward modeling framework, \texttt{PCAT-DE}, to simultaneously constrain the signal components. This enables us to make more efficient use of the available information in the observed data while robustly capturing correlated, non-Gaussian uncertainties on \dIb. We propagate samples from the \texttt{PCAT-DE} posterior to estimates of \dIb\ and use the resulting empirical PDF to constrain the Comptonization parameters. This was important in our analysis -- we discovered that approximating the SPIRE posterior with a Gaussian covariance matrix significantly biased the final constraints.  
    
    As noted in Section~\ref{sS:SZlike}, our constraints on \Ttwofive\ are strongly influenced by the relatively low measured value of \dIplw. A large positive correlation between the measured \dIpmw\ and \dIplw\ values, coupled with them falling on opposite sides of the SZ effect spectrum for intermediate temperatures, results in a bimodal posterior for \Ttwofive\ and a significant expansion of the 68~per cent credible interval range (22.6~keV compared to 14.7~keV when a more likely value of \dIplw\ is assumed). We have not been able to identify any significant biases in the measured \dIplw, but it is possible that one or more do exist. For instance, the particular spatial distribution of astrophysical signals toward \rxj\ may result in PCAT incorrectly associating some of the PLW SZ effect signal with CIB sources. It is also possible that some unmodeled astrophysical components are significant enough to produce a bias, such as the excess SZ effect signal associated with an ICM shock SE of the cluster center \citep{Komatsu2001,Mason2010,Kitayama2016,DiMascolo2019}. \rxj\ also has a pronounced cool core, giving rise to large temperature variations as a function of radius within the ICM. This invalidates the assumption of isothermality along the line of sight, and causes a slight change in the SZ spatial structure between the Bolocam and SPIRE observing bands. Since we are utilizing Bolocam for the SZ effect spatial template, this could change the location and significance of the actual SZ effect ``peak". Another possibility is that the unexplained cluster-coincident emission with a dust-like SED noted by \citet{Erler18} has not been fully described by the combination of the PCAT source catalog and diffuse Fourier component model. 

    Applying our analysis method to a larger sample of galaxy clusters will allow us to better assess if the somewhat low measured value of \dIplw\ is due to a statistical fluctuation, something particular to \rxj, or a more general bias in our methodology. Multi-probe SPIRE, Bolocam, {\it HST}, {\it Chandra}, and {\it Planck} observations exist for an additional 20 galaxy clusters, and another 18 objects can be studied using data from the ACT survey in place of Bolocam \citep{Aiola2020}. In addition to enabling checks for subtle, emergent systematics, this sample of 39 galaxy clusters will allow us to place tight constraints on the ensemble-average temperature measured using the rSZ effect. Taking the raw sensitivity from this study in the absence of a bimodal \Ttwofive\ posterior, we estimate that the sample mean temperature will be measured to a precision of approximately 1~keV. By comparing with pressure-weighted X-ray temperatures, an empirical temperature calibration of {\it Chandra} and/or XMM-{\it Newton} should thus be possible. This level of sensitivity would also allow us to probe the expected differences between rSZ effect and X-ray emission-weighted temperatures due to spatial and thermal inhomogeneities in the ICM \citep{Lee2020}.
    
    Looking forward, potential future wide-field and high angular resolution facilities such as AtLAST and CSST will provide better decoupling of the SZ effect from unwanted astrophysical signals, resulting in higher signal-to-noise ratio rSZ effect measurements of the ICM temperature while probing smaller physical scales. These data would complement future X-ray facilities and open exciting new areas of study, such as routine temperature measurements in the highest redshift galaxy clusters and resolved imaging of the hottest shock-heated gas produced in major merger events. In this work, we have demonstrated an analysis framework that can help maximize the science reach of these facilities.

\section*{Acknowledgments}
    The scientific results reported in this article were based on data obtained from the HerMES project (\url{http://hermes.sussex.ac.uk/}), a Herschel Key Programme utilising Guaranteed Time from the SPIRE instrument team, ESAC scientists and a mission scientist. The HerMES data was accessed through the Herschel Database in Marseille (\url{http://hedam.lam.fr}) operated by CeSAM and hosted by the Laboratoire d'Astrophysique de Marseille. The Bolocam data are publicly available at \url{https://irsa.ipac.caltech.edu/data/Planck/release_2/ancillary-data/bolocam/bolocam.html}. 
    
    The authors acknowledge Research Computing at the Rochester Institute of Technology for providing computational resources and support that have contributed to the research results reported in this publication.
    
    The authors would also like to acknowledge Glenn Morris (\url{https://orcid.org/0000-0003-2985-9962}) for their assistance with \textit{Chandra} data reduction. 
    
   This work was supported by NASA ADAP research grant 80NSSC19K1018. AZ acknowledges support from the Ministry of Science and Technology, Israel.
    


\vspace{5mm}
\facilities{Herschel, HST, CXO, Planck, CSO}

\software{\texttt{astropy} \citep{2013A&A...558A..33A,2018AJ....156..123A}, \texttt{corner} \citep{corner}, \texttt{SZpack} \citep{Chluba2012,Chluba2013}, \texttt{CIAO} (\url{https://cxc.cfa.harvard.edu/ciao/}, Fruscione et al. 2006, SPIE Proc. 6270, 62701V, D.R. Silvia \& R.E. Doxsey, eds.) \texttt{PCAT-DE} (Feder et \emph{al}. 2021, in prep.)}

\bibliography{references}
\bibliographystyle{aasjournal}

\end{document}